\documentclass[a4paper,fleqn]{cas-sc}

\usepackage[authoryear,longnamesfirst]{natbib}
\usepackage{graphicx}
\usepackage{amsmath}
\usepackage{amssymb}
\usepackage{color}
\usepackage{url}
\usepackage{algorithm}
\usepackage{algorithmic}
\usepackage{bm}
\usepackage{cases}
\usepackage{multirow}
\usepackage{multicol}
\usepackage{lineno}

\newcommand{\argmin}{\mathop{\rm arg~min}\limits}

\begin{document}
\let\WriteBookmarks\relax
\def\floatpagepagefraction{1}
\def\textpagefraction{.001}

\shorttitle{Functional-prior-based Bayesian PDE-constrained inversion using PINNs}
\shortauthors{Agata and Okazaki}

\title[mode=title]{Functional-prior-based approaches to Bayesian PDE-constrained inversion using physics-informed neural networks}

\author[1]{Ryoichiro Agata}
\author[2,3]{Tomohisa Okazaki}

\affiliation[1]{organization={Japan Agency for Marine-Earth Science and Technology},
            country={Japan}}
\affiliation[2]{organization={Disaster Prevention Research Institute, Kyoto University},
            country={Japan}}
\affiliation[3]{organization={RIKEN Center for Advanced Intelligence Project},
            country={Japan}}

\begin{abstract}
Physics-informed neural networks (PINNs) provide a mesh-free framework for solving PDE-constrained inverse problems, but their extension to Bayesian inversion still faces a fundamental difficulty: prior distributions are typically defined in the weight space of neural networks, whereas physically meaningful prior assumptions are more naturally expressed in function space. In this study, we introduce a unified framework, termed functional-prior-based approaches to Bayesian PDE-constrained inversion using physics-informed neural networks (fpBPINN), to incorporate functional priors into Bayesian PINN-based inversion. We consider two complementary approaches. The first is a functional-prior-informed Bayesian PINN (FPI-BPINN), in which a neural network weight prior is learned to be consistent with a prescribed functional prior, and Bayesian inference is subsequently performed in weight space. The second is function-space particle-based variational inference for PINNs (fParVI-PINN), which performs Bayesian estimation using ParVI directly in function space. We also show that random Fourier features (RFF) play an important role in representing Gaussian functional priors with neural networks and in improving posterior approximation. We applied the proposed approaches to one-dimensional seismic traveltime tomography and two-dimensional Darcy-flow permeability inversion. These numerical experiments showed that both approaches accurately estimated posterior distributions, highlighting the significance of introducing physically interpretable functional priors into Bayesian PINN-based inverse problems. We also identified the contrasting advantages of FPI-BPINN and fParVI-PINN, namely flexibility and accuracy, respectively.

\end{abstract}

\begin{highlights}
\item Functional priors provide a unified viewpoint for Bayesian PDE-constrained inversion with PINNs.
\item FPI-BPINN and fParVI-PINN enable uncertainty quantification with physically interpretable priors.
\item Random Fourier features improve Gaussian-prior representation and posterior approximation.
\end{highlights}


\maketitle

\section{Introduction}

Inverse problems that estimate unknown parameters from observational data while enforcing the governing partial differential equations (PDEs) to be satisfied, which we refer to as PDE-constrained inverse problems, play a central role in many fields of science and engineering, including geoscience, fluid mechanics, materials science, and thermal engineering. These problems are generally nonlinear and ill-posed and are often strongly affected by observational noise and model uncertainty. Therefore, in addition to deterministic estimation based on discretized forward solvers and optimization, it is important to adopt a Bayesian framework that quantifies uncertainty. Traditionally, forward analyses, adjoint methods, and sampling methods based on grid- or mesh-based representation have been the mainstream Bayesian approaches.

However, in recent years, mesh-free inverse analysis using deep learning techniques has attracted attention as an alternative approach. A representative example is the physics-informed neural network (PINN) \cite[]{Raissi2019}, which provides a framework for treating both forward and PDE-constrained inverse problems in a unified manner by incorporating the residual of the governing equation into the loss function. A major advantage of PINNs in inverse problems is that not only the solution of the governing equation but also unknown quantities such as coefficient fields and source terms can be represented by neural networks (NNs), enabling estimation in a fully mesh-free manner. On the other hand, standard PINNs are essentially point-estimation methods and cannot directly handle uncertainty arising from observational noise or data sparseness. 
Introducing uncertainty quantification (UQ) based on Bayesian inference into PINN-based inverse analyses is therefore a promising direction. Compared with other approaches such as stochastic-process-based regression methods, PINN-based Bayesian approaches can more naturally incorporate PDE constraints, including those involving nonlinear differential operators, while retaining the scalability of NN-based training with respect to the number of data points. However, such efforts are still at an early stage. A pioneering example is Bayesian physics-informed neural networks (B-PINNs) \cite[]{Yang2021}, in which Bayesian neural networks (BNNs) are incorporated into PINNs, providing a framework for estimating posterior distributions using Hamiltonian Monte Carlo (HMC) \cite[]{Duane1987} and Variational Inference (VI) \cite[]{Jordan1999ML}. Similar problems have also been solved using different Bayesian methods \cite[]{Sun2020TAML,Pensoneault2024JCP}, such as particle-based variational inference (ParVI) \cite[]{Liu2016,Liu2019ICML} and ensemble Kalman inversion (EnKI) \cite[]{Kovachki2019IP}. B-PINNs demonstrated that forward and PDE-constrained inverse problems with noisy data can be treated in a Bayesian manner.
Unlike ordinary Bayesian inference, in which unknown parameters are estimated directly, unknown parameters in BNNs are usually the weights of the NNs. In BNNs, the prior distribution required for Bayesian inference is specified on the NN weights. Typically, simple priors, such as independent and identically distributed (IID) Gaussian distributions or Student's t-distributions, are assigned to the weights. However, such simple priors in weight space are difficult to interpret physically in the function space represented by the NN outputs and may introduce uncontrolled effects into the inference results \cite[]{Wang2019}.

A fundamental issue here is therefore the design of prior distributions when PDE parameters are represented by NNs. \cite{Tran2022} identified this as a fundamental limitation of BNNs and proposed a framework for training NN weight priors to match a desired functional prior defined by a Gaussian process (GP) through Wasserstein distance minimization. In other words, function space is more physically and statistically interpretable than weight space, and this viewpoint is also important for PDE-constrained inverse problems. Recent studies have continued to propose the design of weight priors \cite[]{Alberts2026CMAME} and activation functions \cite[]{Sendera2025UAI} such that the stochastic behavior of NN output functions more closely resembles that of favorable GPs in function space.  The idea of focusing on priors in function space in BNN has also attracted attention in the context of Generative Adversarial Networks (GANs) \cite[]{Goodfellow2014NeurIPS,Meng2022JCP}, diffusion models \cite[]{Ho2020NeurIPS,Li2025arXiv}, and anchored ensembling \cite[]{Ghorbanian2025CMAME}.
However, in the context of learning weight priors that are consistent with a functional prior for PINN-based inverse problems, this issue has not yet been investigated.
Related to this line of research, studies that formulate UQ for BNNs directly in function space have also been developed \cite[]{Sun2019,Wang2019,Rudner2022NeurIPS}. \cite{Wang2019} proposed function-space ParVI (fParVI), which performs ParVI in function space rather than in weight space, thereby avoiding the degeneration of particle-based methods in overparameterized models. Since this method formulates Bayesian estimation in function space, a functional prior in closed form can be directly incorporated. This approach was also interpreted by \cite{DAngelo2021NeurIPS} as a deep ensemble with a repulsive term in function space. These approaches that perform Bayesian inference directly in function space were, to the best of our knowledge, first applied to PINN-based inverse problems in the context of solid-Earth geophysics by \cite{Agata2023}. Similar ideas have since been proposed independently by \cite{Rover2024arXiv,Pilar2025arXiv,Shukla2026arXiv}, although introducing appropriate priors is not their main focus.

In this study, we therefore introduce a unified framework, which we call functional-prior-based approaches to Bayesian PDE-constrained inversion using physics-informed neural networks (fpBPINN), for Bayesian inverse problems of PDE parameters based on PINNs. fpBPINN focuses on priors that have a clear meaning in the function space represented by NN outputs. Specifically, we compare two approaches: a functional-prior-informed Bayesian PINN (FPI-BPINN), which first learns an NN weight prior consistent with a prescribed functional prior and then performs Bayesian inference in weight space, and fParVI for PINNs (fParVI-PINN), which performs ParVI for PDE parameters directly in function space. The former offers the flexibility of exploiting existing BNN inference methods after prior learning, whereas the latter allows prior distributions to be described naturally in function space. Furthermore, we point out that, when applying these methods to PINN-based inverse analyses, the introduction of random Fourier features (RFF) into the NN plays an important role in facilitating the learning of frequency components corresponding to the prior distribution in function space. Through one-dimensional (1D) seismic traveltime velocity estimation, we demonstrate the importance of considering a functional prior when aiming for Bayesian estimation consistent with physical intuition and show that both proposed approaches achieve accurate functional-prior-based inference. In two-dimensional (2D) Darcy-flow permeability estimation, we validate the applicability of the proposed methods to more realistic problem settings. We clarify the significance of introducing physically interpretable functional priors into PINN-based inverse problems and compare the advantages of the two approaches.

The remainder of this paper is organized as follows. Section 2 formulates the Bayesian PDE-constrained inversion problem and summarizes the adjoint-based gradient evaluation used in the proposed framework. Section 3 introduces the functional-prior-based Bayesian PINN approaches, including FPI-BPINN and fParVI-PINN. Section 4 presents numerical experiments on 1D seismic traveltime tomography and 2D Darcy-flow permeability inversion. Section 5 discusses the implications, advantages, and limitations of the proposed approaches. Finally, Section 6 provides a conclusion.

\section{Formulation of Bayesian inversion problem}
\subsection{Neural network-based formulation of PDE-constrained inversion}

Let $\Omega \subset \mathbb{R}^d$ be a spatial domain with boundary $\partial\Omega$.
A general PDE constraint can be formulated in operator form as
\begin{eqnarray}
\mathcal{F}(u(\textbf{x}),m(\textbf{x}))=0
\qquad \textbf{x} \in \Omega, \label{eq:pde_constraint}\\
\mathcal{B}(u(\textbf{x})) = 0
\qquad \textbf{x} \in \partial\Omega, \label{eq:boundary_condition}
\end{eqnarray}
where $\mathcal{F}$ and $\mathcal{B}$ represent the (possibly nonlinear) differential operators defining the physics and boundary conditions, respectively.
$u(\textbf{x})$ is the PDE solution at point $\textbf{x}$.
We define a PDE-constrained inverse problem as the problem of estimating the parameter field $m(\textbf{x})$ that satisfies the PDE constraints (Equations \ref{eq:pde_constraint} and \ref{eq:boundary_condition}) when the data $\mathcal{D}$ are given as
\begin{equation}
\mathcal{D} = \{(x_i,u(\textbf{x}_i))\}_{i=1}^{N_d},\textbf{x}_i \in \Omega,
\end{equation}
where $\textbf{x}_i$ are the observation points.
The deterministic inverse problem is formulated as the problem of obtaining the parameter field $m(\textbf{x})$ that minimizes both the misfit between the observed and predicted data and the residual of the PDE constraint.
The conventional approach to this problem is to represent the solution $u(\textbf{x})$ and the parameter field $m(\textbf{x})$ by discretizing the domain $\Omega$ using a grid or mesh.
However, the emergence of PINNs has enabled such inverse problems to be solved in a mesh-free manner, representing the solution and the parameter field using NNs as
\begin{eqnarray}
    u(\textbf{x}) \approx f_u(\textbf{x},{\boldsymbol{\theta}_u}),
\qquad \textbf{x} \in \Omega, \\
m(\textbf{x}) \approx f_m(\textbf{x},{\boldsymbol{\theta}_m}),
\qquad \textbf{x} \in \Omega,
\end{eqnarray}
where $f_u$ and $f_m$ are the NNs representing the solution and the parameter field, respectively.
${\boldsymbol{\theta}_u}$ and ${\boldsymbol{\theta}_m}$ are the weight parameters of the NNs.

\subsection{Bayesian formulation and calculation of gradient of posterior distribution}
\label{sec:adjoint_bayesian}

In the NN-based formulation, Bayes' theorem is generally formulated for the weight parameters ${\boldsymbol{\theta}_m}$ rather than the parameter field $m$ itself, following the standard BNN framework, as
\begin{equation}
p({\boldsymbol{\theta}_m} | \mathcal{D})
=
\frac{p(\mathcal{D}| {\boldsymbol{\theta}_m})\,p({\boldsymbol{\theta}_m})}{p(\mathcal{D})}
\propto
p(\mathcal{D}| {\boldsymbol{\theta}_m})\,p({\boldsymbol{\theta}_m}),
\label{eq:bayes}
\end{equation}
where $p({\boldsymbol{\theta}_m})$ is the prior distribution, $p(\mathcal{D}| {\boldsymbol{\theta}_m})$ is the likelihood, and $p(\mathcal{D}) = \int p(\mathcal{D}| {\boldsymbol{\theta}_m})\,p({\boldsymbol{\theta}_m})\,d{\boldsymbol{\theta}_m}$ is the evidence (marginal likelihood).
The probability distribution function (PDF) for $m(\textbf{x})$, which we actually want to obtain, is obtained as the predictive distribution
\begin{equation}
p(m(\textbf{x}) | \mathcal{D})
=
\int p(m(\textbf{x}) | {\boldsymbol{\theta}_m})\,p({\boldsymbol{\theta}_m} | \mathcal{D})\,d{\boldsymbol{\theta}_m}
\end{equation}
where $p(m(\textbf{x}) | {\boldsymbol{\theta}_m})$ is given by the forward propagation of the NN.
We perform Bayesian estimation of $p({\boldsymbol{\theta}_m} | \mathcal{D})$ in a form that enables integration (marginalization) over ${\boldsymbol{\theta}_m}$, e.g., Monte Carlo integration with particle approximation, analytical integration with Gaussian approximation, or related approximations.
In this paper, we focus on particle approximation; i.e., $p({\boldsymbol{\theta}_m} | \mathcal{D})$ is estimated in the form of ensemble modeling.
Efficient Bayesian inference methods, such as HMC and ParVI, leverage $-\nabla_{\boldsymbol{\theta}_m} \log p(\boldsymbol{\theta}_m | \mathcal{D})$, the gradient of the negative log-posterior PDF with respect to the estimated parameters.
This gradient must be evaluated subject to the PDE constraints (Equations \ref{eq:pde_constraint} and \ref{eq:boundary_condition}).

Thus, we compute the gradient $-\nabla_{\boldsymbol{\theta}_m} \log p(\boldsymbol{\theta}_m | \mathcal{D})$ under the PDE constraints using the Lagrange multiplier method, also known as the adjoint method.
Let $\mathcal{J}$ denote the loss function, with $\mathcal{J}=-\log p(\boldsymbol{\theta}_m | \mathcal{D})$.
To incorporate the PDE and boundary constraints, we introduce the continuous Lagrangian functional
\begin{equation}
\mathcal{L}_c(\boldsymbol{\theta}_u,\boldsymbol{\theta}_m,\lambda_{\mathcal{F}},\lambda_{\mathcal{B}})
=
\mathcal{J}(\boldsymbol{\theta}_u,\boldsymbol{\theta}_m)
+
\left\langle \lambda_{\mathcal{F}},\,\mathcal{F}\bigl(f_u(\boldsymbol{\theta}_u),f_m(\boldsymbol{\theta}_m)\bigr)\right\rangle_\Omega
+
\left\langle \lambda_{\mathcal{B}},\,\mathcal{B}\bigl(f_u(\boldsymbol{\theta}_u)\bigr)\right\rangle_{\partial\Omega},
\label{eq:lagrangian_functional}
\end{equation}
where $\lambda_{\mathcal{F}}$ and $\lambda_{\mathcal{B}}$ are the adjoint variables associated with the PDE and boundary constraints, respectively.
Here, $\langle\cdot,\cdot\rangle_\Omega$ and $\langle\cdot,\cdot\rangle_{\partial\Omega}$ denote appropriate duality pairings over $\Omega$ and $\partial\Omega$.
For notational simplicity, the spatial dependence of $f_u(\boldsymbol{\theta}_u)$ and $f_m(\boldsymbol{\theta}_m)$ is omitted.
Taking the variation of $\mathcal{L}_c$ with respect to $\boldsymbol{\theta}_u$ and $\boldsymbol{\theta}_m$, we obtain the adjoint equations for $\lambda_{\mathcal{F}}$ and $\lambda_{\mathcal{B}}$.
Subsequently, $\nabla_{\boldsymbol{\theta}_m} \mathcal{J}$ can be calculated using the solution of the adjoint equations.
See Appendix \ref{sec:appendix_adjoint} for details of the derivation.

In conducting Bayesian estimation, we iteratively evaluate the gradient (Equation \ref{eq:gradient_of_loss_function}) and update the weight parameters ${\boldsymbol{\theta}_m}$.
By construction in the adjoint method, the PDE constraints $\mathcal{F}\bigl(f_u(\boldsymbol{\theta}_u),f_m(\boldsymbol{\theta}_m)\bigr)=0$ and $\mathcal{B}\bigl(f_u(\boldsymbol{\theta}_u)\bigr)=0$ should be satisfied within numerical error at every iteration.
We conduct PINN training to minimize the loss for the PDE constraints (Equations \ref{eq:pde_constraint} and \ref{eq:boundary_condition}) for the current weight parameters ${\boldsymbol{\theta}_m}^{\rm current}$, as
\begin{eqnarray}
    \label{eq:pinn_training}
    {\boldsymbol{\theta}_u} = \argmin_{{\boldsymbol{\theta}_u}} \mathcal{L}_{\rm PDE}({\boldsymbol{\theta}_u}, {\boldsymbol{\theta}_m}^{\rm current}) + \mathcal{L}_{\rm boundary}({\boldsymbol{\theta}_u})
\end{eqnarray}
where
\begin{eqnarray}
    \mathcal{L}_{\rm PDE}({\boldsymbol{\theta}_u}, {\boldsymbol{\theta}_m}^{\rm current}) \simeq \int_\Omega \left( \mathcal{F}(u,m) \right)^2 d\Omega \\
    \mathcal{L}_{\rm boundary}({\boldsymbol{\theta}_u}) \simeq \int_{\partial\Omega} \left( \mathcal{B}(u) \right)^2 d\partial\Omega.
\end{eqnarray}
These integrations are approximated by Monte Carlo integration using collocation points randomly generated in $\Omega$ and on $\partial\Omega$.
Because the weight parameters obtained in the previous iterative step are already close to the optimal values for the current step, a relatively small number of epochs is required for PINN training in each step.

This algorithm separates the iterative update, i.e., Bayesian estimation, for ${\boldsymbol{\theta}_m}$ from the PINN training for ${\boldsymbol{\theta}_u}$.
As a result, only the parameter-field learning is treated in a Bayesian manner, whereas the PINN representing the PDE solution is trained deterministically.
This contrasts with commonly used PINN-based inversion methods, where ${\boldsymbol{\theta}_m}$ and ${\boldsymbol{\theta}_u}$ are simultaneously trained based on a single loss function defined as
\begin{eqnarray}
    \mathcal{L}_{\rm total} = \mathcal{L}_{\rm PDE} + \mathcal{L}_{\rm boundary} + \mathcal{J},
\end{eqnarray}
where the negative log-posterior $\mathcal{J}$ corresponds to data misfit in this context.
If we extend such a simultaneous approach to Bayesian estimation, the target posterior PDF would be $p({\boldsymbol{\theta}_m}, {\boldsymbol{\theta}_u} | \mathcal{D})$.
However, adding ${\boldsymbol{\theta}_u}$ to the Bayesian estimation would substantially increase the dimension of the parameter space and introduce scale differences, which may lead to severe difficulties in the estimation.

\section{Function-space approaches to PINN-based Bayesian PDE-constrained inversion}
\subsection{Approach 1: Functional-prior-informed Bayesian physics-informed neural network (FPI-BPINN)}
\label{sec:FPI-BPINN}

In BNNs, the prior distribution of the weight parameters ${\boldsymbol{\theta}_m}$ is typically set to a simple distribution such as an independent and identically distributed (IID) normal distribution. 
However, such a prior is known to be difficult to interpret in the function space of the NN and often leads to uncontrolled adverse effects in the estimation results \cite[]{Wang2019,Matsubara2021JMLR}.
Instead, we define a plausible stochastic process in the function space of the NN and train $p({\boldsymbol \theta}_m)$ based on it. 
The stochastic behavior of the output function of the trained NN is expected to become similar to that of the prescribed function-space stochastic process. 
This follows the approach proposed in the general BNN context by \cite{Tran2022}.
We apply this learned prior PDF to PINN-based Bayesian PDE-constrained inversion. We term this approach FPI-BPINN.

We assume an independent normal distribution for the prior PDF of the weight parameters ${\boldsymbol \theta}_m$, as
\begin{eqnarray}
p({\boldsymbol \theta}_m)=\mathcal{N}(\boldsymbol{\mu},\text{diag}(\boldsymbol{\sigma}^2)),
\end{eqnarray}
where $\boldsymbol{\mu}$ and $\boldsymbol{\sigma}$ are the mean and standard deviation of the prior PDF, respectively.
Both have the same dimension as the number of weight parameters ${\boldsymbol \theta}_m$.
$\text{diag}(\cdot)$ is a diagonal matrix whose diagonal elements are given by the input vector.
Let $\mathcal{SP}(\psi)$ denote the target stochastic process defined in the function space of the NN and parameterized by $\psi$.
We learn $\boldsymbol{\mu}$ and $\boldsymbol{\sigma}$ so that $p({\boldsymbol \theta}_m)$ behaves similarly to $\mathcal{SP}(\psi)$.
In most cases, a Gaussian process is the first candidate for the target stochastic process, i.e., $\mathcal{SP}(\psi)$ = $\mathcal{GP}(\mu_{\mathcal{GP}}(\textbf{x}), k_{\mathcal{GP}}(\textbf{x},\textbf{x}'))$, where $\mu_{\mathcal{GP}}(\textbf{x})$ and $k_{\mathcal{GP}}(\textbf{x},\textbf{x}')$ are the mean function and kernel function, respectively.
A widely used kernel function is the radial basis function (RBF) kernel, which we define as
\begin{eqnarray}
    k_{\mathcal{GP}}(\textbf{x}, \textbf{x}') = a^2 \exp\left(-\frac{1}{l_\mathcal{GP}^2} \|\textbf{x} - \textbf{x}'\|^2\right),
\end{eqnarray}
where $l_\mathcal{GP}$ is the correlation length scale and $a$ is the amplitude of the kernel function, corresponding to the standard deviation of the marginal probability.

\cite{Tran2022} proposed using the Wasserstein distance, which involves nested optimization by solving an optimal transport problem in each evaluation, as a metric for measuring the discrepancy between two probability distributions to construct the loss function for learning. Instead, we adopt the maximum mean discrepancy (MMD), which can be estimated directly from samples using pairwise kernel evaluations, primarily for computational and implementation simplicity.
MMD is calculated for random samples taken from $p({\boldsymbol \theta}_m)$ and $\mathcal{SP}(\psi)$, as
\begin{eqnarray}
    \mathcal{L}_{\rm MMD} = \frac{1}{N^2} \sum_{i=1}^N \sum_{j=1}^N k_{\rm MMD}(f_m^{(i)}, f_m^{(j)}) - \frac{2}{N^2} \sum_{i=1}^N \sum_{j=1}^N k_{\rm MMD}(f_m^{(i)}, m_\mathcal{SP}^{(j)}) + \frac{1}{N^2} \sum_{i=1}^N \sum_{j=1}^N k_{\rm MMD}(m_\mathcal{SP}^{(i)}, m_\mathcal{SP}^{(j)}),
    \label{eq:mmd_loss}
\end{eqnarray}
where $\{f_m^{(i)}\}_{i=1}^N$ and $\{m_\mathcal{SP}^{(j)}\}_{j=1}^N$ denote the sets of random samples drawn from $p(\theta_m)$ and $\mathcal{SP}(\psi)$, respectively.
Here, $k_{\rm MMD}(\cdot, \cdot)$ represents a positive definite kernel function, for which we use the RBF kernel in this study.
The reader should be careful not to confuse $k_{\rm MMD}(\cdot, \cdot)$ with the kernel function $k_{\mathcal{GP}}(\cdot, \cdot)$ used for the Gaussian process.
The disadvantage of MMD compared with the Wasserstein distance is that the bandwidth of the RBF kernel is introduced as a hyperparameter.
Determining it using the median heuristic works well in our cases.
We obtain $\boldsymbol{\mu}$ and $\boldsymbol{\sigma}$ that minimize the loss function by using a stochastic gradient descent algorithm.
After learning, the prior PDF $p(\theta_m)$ with the learned $\boldsymbol{\mu}$ and $\boldsymbol{\sigma}$ is incorporated into the Bayesian estimation of the weight parameters $\theta_m$. 
In summary, FPI-BPINN imposes the PDE constraint on the parameter-field BNN, whose prior is trained for a favorable functional GP with MMD-based loss, through adjoint-based gradient calculation with PINN-based forward modeling described in Section \ref{sec:adjoint_bayesian}.

FPI-BPINN can be applied to any Bayesian estimation method that uses the gradient of the posterior PDF.
However, each evaluation of the gradient of the log posterior PDF is computationally expensive because it requires one adjoint computation and one PINN training step.
Thus, in practice, application to a Bayesian estimation method that requires many sequential evaluations of the gradient of the loss function, such as the Hamiltonian Monte Carlo (HMC) method \cite[]{Duane1987}, may not be promising.
We use Stochastic Gradient Langevin Dynamics (SGLD) plus a repulsion force (SGLD+R) \cite[]{Gallego2018NeurIPS} to perform Bayesian estimation.
SGLD+R is a variant of SGLD, a widely used stochastic gradient Markov chain Monte Carlo (SG-MCMC) method, which introduces multiple chains and uses a repulsion force between the chains to prevent the particles from clustering.
From the viewpoint of ParVI, SGLD+R is interpreted as a variant of Stein variational gradient descent (SVGD) \cite[]{Liu2016} that adds a noise term to the update equation.
The combination of a stochastic gradient and a repulsion force makes SGLD+R a highly parallelizable Bayesian estimation method, requiring a moderate number of sequential evaluations, typically on the order of $10^3$, of the gradient of the loss function.
We apply a preconditioning technique to the update equation of SGLD+R to improve the convergence speed (see Section \ref{sec:preconditioning_sgldr}).

We summarize this two-stage FPI-BPINN procedure in Algorithm \ref{alg:fpi_bpinn}. 

\subsection{Approach 2: Function-space particle-based variational inference for PINN (fParVI-PINN)}

The second approach, which we term fParVI-PINN, formulates and performs Bayesian estimation directly for the NN output values at selected evaluation points in function space, as
\begin{eqnarray}
p(\textbf{m} | \mathcal{D}) =
\frac{p(\mathcal{D}| \textbf{m})\,p(\textbf{m})}{p(\mathcal{D})}
\propto
p(\mathcal{D}| \textbf{m})\,p(\textbf{m}),
\label{eq:bayes_function_space}
\end{eqnarray}
where $\textbf{m} = \{f_m(\mathbf{x}_i)\}_{i=1}^{N_d}$ denotes the output values of the NN at $N_d$ evaluation points.
In fact, this equation appears similar to ordinary Bayesian estimation with a grid-based representation of the parameter field.
fParVI \cite[]{Wang2019} performs Bayesian estimation defined in function space, while using an NN-based parameterization.
fParVI calculates the update vector of ParVI in function space and converts it to the weight space using the Jacobian matrix of the NN.
In the case of SVGD, the most widely used instance of ParVI, the update vector for fParVI is given as
\begin{eqnarray}
\boldsymbol{\theta}_{m\,i}^{l+1}=\boldsymbol{\theta}_{m\,i}^{l}+\epsilon_{l}
\frac{\partial {\bf m}_{i}}{\partial \boldsymbol{\theta}_{m\,i}^{l}}^{\top}  \boldsymbol{\phi}\left(\textbf{m}_{i}^{l+1}\right).
\label{eqn:SVGD_update_improved2}\\
\boldsymbol{\phi}(\mathbf{m})=\frac{1}{n} \sum_{j=1}^{n} \lbrace k(\mathbf{m}_{j}^{l},\mathbf{m}) \nabla_{\mathbf{m}_{j}^{l}}\log P(\mathbf{m}_{j}^{l}|\mathcal{D})+\nabla_{\mathbf{m}_{j}^{l}} k(\mathbf{m}_{j}^{l},\mathbf{m}) \rbrace,
\label{eqn:SVGD_vector}
\end{eqnarray}
This algorithm avoids introducing $p(\boldsymbol{\theta}_m)$ into the Bayesian estimation but instead directly incorporates the prior PDF in function space, $p(\textbf{m})$.
For example, we can apply a Gaussian process, for which the mean and covariance are far more controllable and physically interpretable, to $p(\textbf{m})$.
$\nabla_{\mathbf{m}_{j}^{l}}\log P(\mathbf{m}_{j}^{l}|\mathcal{D})$ can be calculated by slightly modifying the adjoint-method formulation given in Section \ref{sec:adjoint_bayesian}.

\begin{figure}[htbp]
    \centering
    \includegraphics[width=0.5\textwidth]{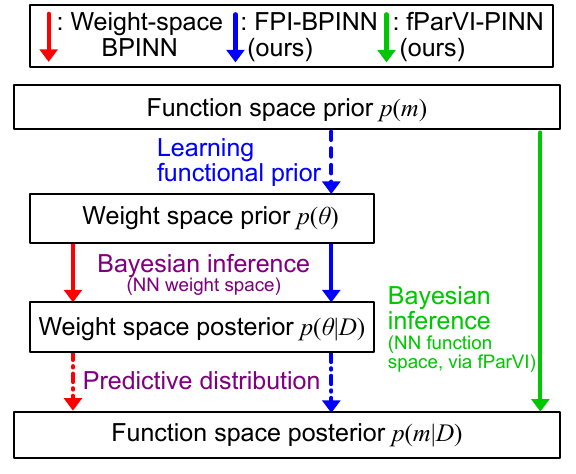}
    \caption{Schematic comparison among the weight-space Bayesian PINNs (BPINN, e.g., \cite{Yang2021}), and the proposed fpBPINN approaches, including FPI-BPINN and fParVI-PINN. The items shown in purple are applied to both BPINN and FPI-BPINN.}
    \label{fig:method_comparison}
\end{figure}

We summarize this learning procedure in Algorithm \ref{alg:fparvi_pinn}.
Although the basic idea of fParVI-PINN was introduced in \cite{Agata2023}, we newly introduce random Fourier features (RFF) with theoretically motivated characteristic frequency parameter to represent a Gaussian process in function space, as shown in the following subsection. This incremental development of fParVI-PINN was strongly inspired by the findings obtained during the development of the algorithm for functional-prior learning in FPI-BPINN.
We also summarize the comparison among the weight-space Bayesian PINNs (e.g., \cite{Yang2021}), FPI-BPINN, and fParVI-PINN in Fig. \ref{fig:method_comparison}.

\subsection{Importance of random Fourier features in representing Gaussian functional priors}
\label{sct:RFF}

We assume the use of a Gaussian process to provide the prior PDF $p({\boldsymbol \theta}_m)$ for FPI-BPINN and $p(\textbf{m})$ for fParVI-PINN.
Both methods require the PDE-parameter NN to represent a Gaussian process in function space.
It has been reported that a simple fully-connected NN (FCNN) has difficulty learning functional Gaussian processes in certain situations, and that introducing a periodic activation function based on trigonometric functions is effective for improving the learning \cite[]{Sendera2025UAI}.
Inspired by their work, we introduce random Fourier features \cite[]{Tancik2020} into the PDE-parameter NN to improve its ability to represent Gaussian processes.
An FCNN with RFF is formally defined as
\begin{eqnarray}
f_{\mathrm{RFF}}(\textbf{x}; \boldsymbol{\theta})
    = f_L \circ f_{L-1} \circ \cdots \circ f_1 \bigl( \gamma(x) \bigr).
\end{eqnarray}
where
\begin{eqnarray}
\gamma(x)
:= \left[\cos(2\pi \textbf{Bx}),\, \sin(2\pi \textbf{Bx})\right],\\
B_{ij} \sim \mathcal{N}(0, \tau^{2}),\\
f_\ell(\mathbf{h}) = \psi\left(\mathbf{W}_\ell \mathbf{h} + \mathbf{b}_\ell\right),
\end{eqnarray}
where $f_\ell(\cdot)$ is the neural network function in the $\ell$-th layer, $\psi(\cdot)$ is the activation function, and $\mathbf{W}_\ell$ and $\mathbf{b}_\ell$ are the weight matrix and bias vector in the $\ell$-th layer, respectively.
$\boldsymbol{\theta} = \{\mathbf{W}_1, \mathbf{b}_1, \mathbf{W}_2, \mathbf{b}_2, \cdots, \mathbf{W}_L, \mathbf{b}_L\}$ denotes the weight parameters of the NN.
$\mathbf{h}$ is the input vector to the $\ell$-th layer.
Although the characteristic frequency $\tau$ is usually treated as a hyperparameter, we set it so that the output function is well adapted to the target Gaussian process.
Such a $\tau$ can be derived by comparing the Gaussian function whose bandwidth is $\tau$ with the Fourier transform of the kernel function of the target Gaussian process \cite[]{Rasmussen2006}.
Specifically, in the case of the RBF kernel, we set it as
\begin{eqnarray}
\tau = \frac{1}{\sqrt{2}\pi l_\mathcal{GP}}. \label{eqn:characteristic_frequency}
\end{eqnarray}
See Appendix \ref{sec:appendix_characteristic_frequency} for details of the derivation. \cite{Sendera2025UAI} proposed to use an activation function represented as a combination of multiple sinusoidal components with different frequencies, whose amplitudes and frequencies are treated as trainable parameters, to train NN for a functional Gaussian process. By contrast, the matrix $\mathbf{B}$ in RFF used in our framework is fixed a priori based on theoretically motivated parameter choice. The motivation to adopt the proposed method is that it reduces trainable degrees of freedom and stabilizes the learning process by providing an explicit link between the prescribed GP functional prior and the periodic components of the NN. 	In addition, the combination of RFF and Mish activation is a more established and conservative architectural choice that relies on widely used components, and is therefore expected to provide high robustness.

\begin{figure*}[htbp]
    \centering
    \includegraphics[width=0.80\textwidth, bb = 0 0 346 536]{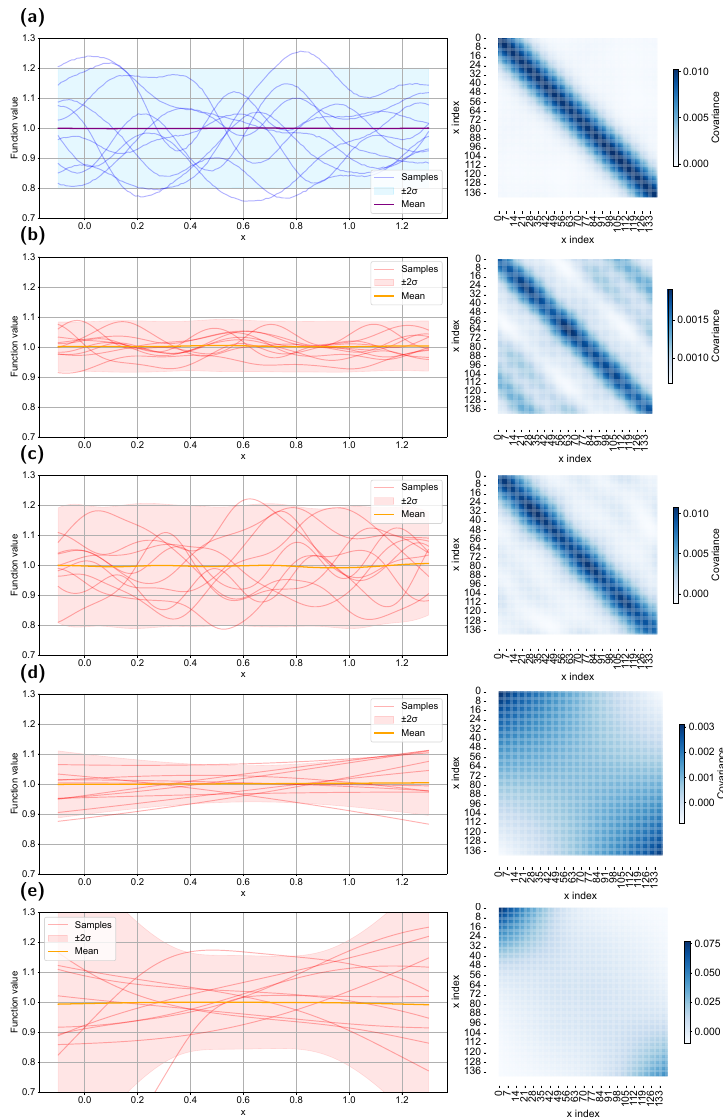}
    \caption{Prior learning results for a one-dimensional Gaussian process with $l_\mathcal{GP}=0.15$. (a) Samples from the target Gaussian process and the true covariance matrix. (b) Samples and sample covariance matrices from a BNN represented by an FCNN with RFF before prior learning. (c) Same as (b), but after prior learning. (d) Samples and sample covariance matrices from a BNN represented by an FCNN without RFF before prior learning. (e) Same as (d), but after prior learning.}
    \label{fig:prior_learning}
\end{figure*}

\begin{figure}[htbp]
    \centering
    \includegraphics[width=0.5\textwidth, bb = 11 11 566 496]{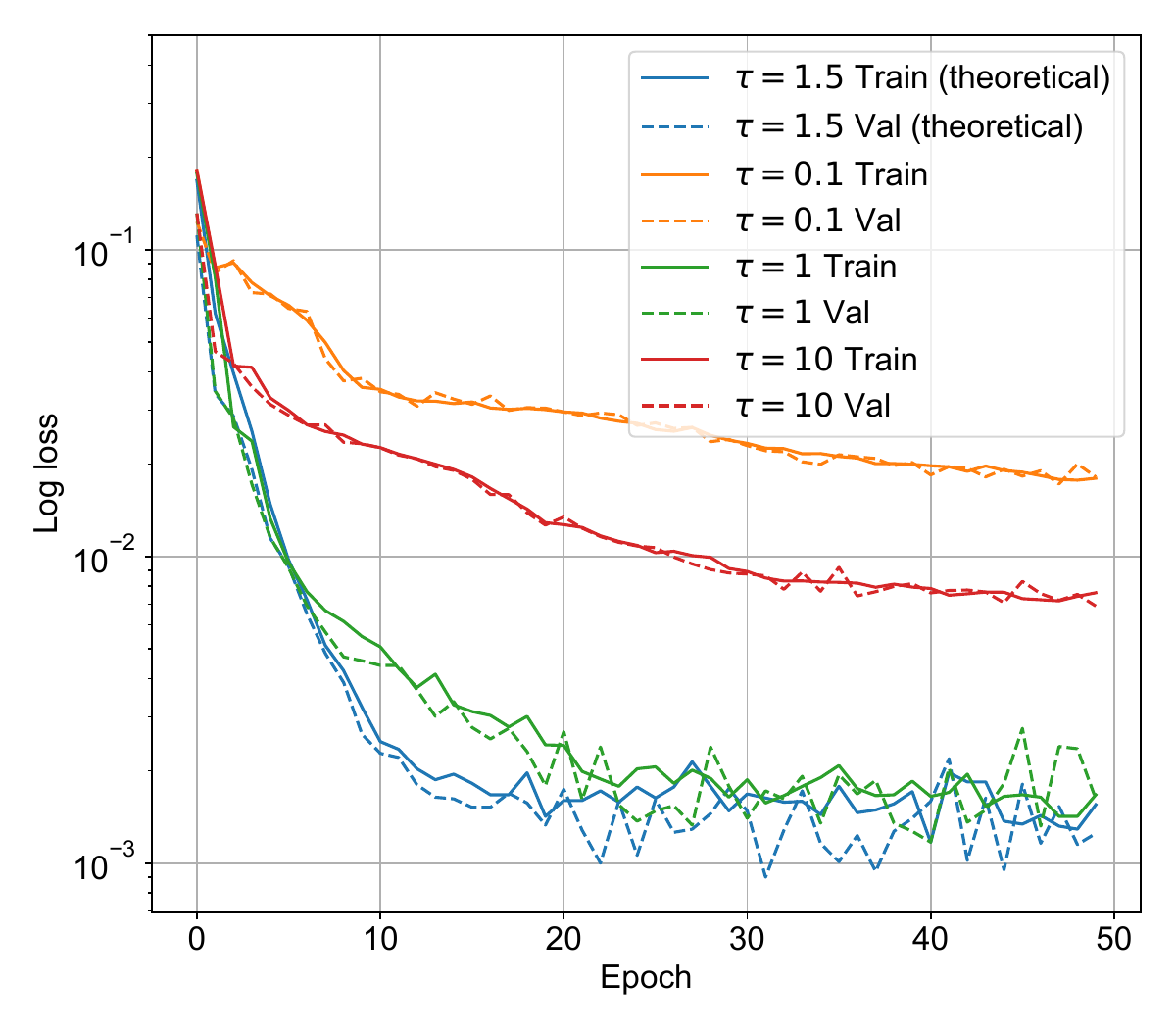}
    \caption{Convergence history of the MMD loss in prior learning for different characteristic frequencies $\tau$ of the RFF for the fixed GP kernel. The theoretically derived value, $\tau=1.5$, provides better convergence than arbitrarily chosen frequencies.}
    \label{fig:loss_history_prior}
\end{figure}

We demonstrate an example of learning a 1D Gaussian process with $l_\mathcal{GP}=0.15$ (Fig. \ref{fig:prior_learning} (a)) using an FCNN with and without RFF.
The FCNN employed two hidden layers and 30 hidden units per layer with the Mish activation function \cite[]{Misra2019arXiv}.
The shape of $\mathbf{B}$ was (15, 1) in the case of RFF. This means that sine and cosine functions with 15 distinct frequencies are introduced into the NN, while the column dimension of 1 corresponds to the dimension of the input $\mathbf{x}$. An excessively small number of rows is expected to reduce the representational power, failing to capture the characteristics of the GP functional prior.
We observed that the output of the FCNN with RFF based on weight samples obtained from the learned mean and standard deviation closely approximated samples from the target Gaussian process after learning.
The sample covariance matrix also showed good agreement with the target covariance matrix (Fig. \ref{fig:prior_learning} (b)).
By contrast, the FCNN without RFF failed to approximate the target Gaussian process well (Fig. \ref{fig:prior_learning} (c)).
The theoretically derived characteristic frequency $\tau=1.5$ yielded consistently better convergence of the MMD-based loss function than cases with arbitrarily chosen $\tau$ for the fixed target GP kernel (Fig. \ref{fig:loss_history_prior}). This theoretically motivated choice of $\tau$ is expected to contribute to stabilizing the functional-prior learning and the subsequent posterior estimation compared to arbitrarily chosen or fully trainable RFF frequency components.
Even in fParVI-PINN, which does not require this learning process, such a proper introduction of RFF is necessary to obtain a good approximation of the posterior PDF when a Gaussian-process prior is used.

\section{Numerical experiments}
\subsection{1D seismic traveltime tomography}

\begin{figure*}[htbp]
    \centering
    \includegraphics[width=0.82\textwidth, bb = 0 0 345 507]{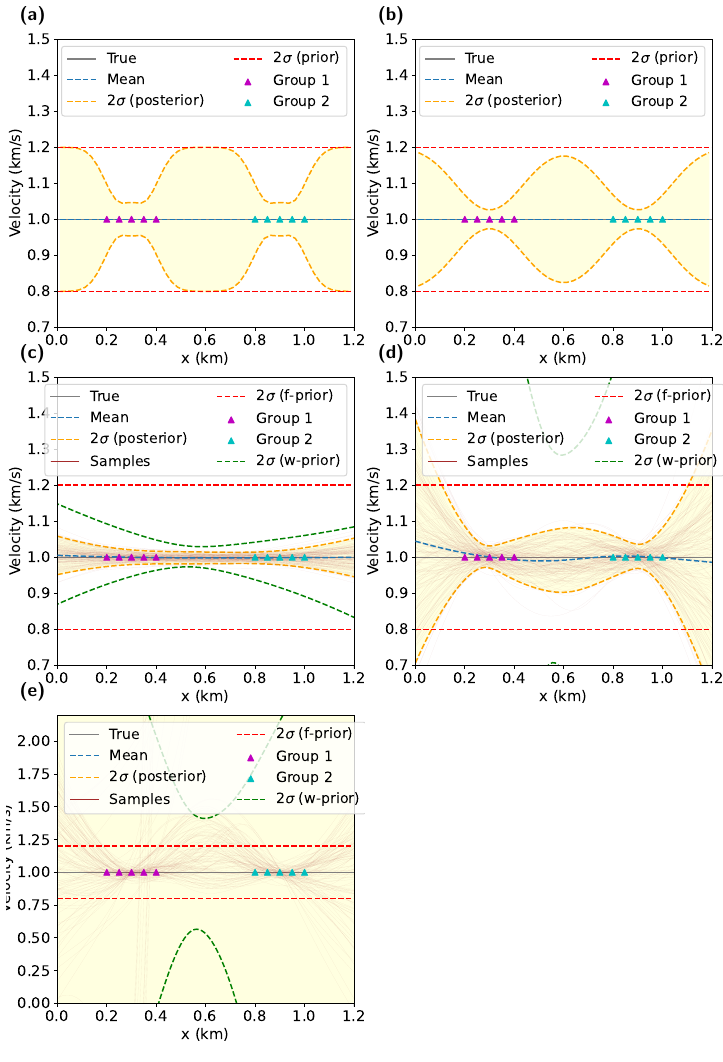}
    \caption{Comparison of posterior velocity distributions for 1D seismic traveltime tomography between the semi-analytical results of functional-prior-based Bayesian estimation and those of weight-space Bayesian estimation using an IID Gaussian prior on an FCNN without RFF. Each panel shows the posterior mean and standard deviation of the velocity. (a) Semi-analytical posterior for $l_\mathcal{GP}=0.075\,{\rm km}$. (b) Same as (a), but for $l_\mathcal{GP}=0.15\,{\rm km}$. (c) Weight-space Bayesian PINN result with $\sigma=0.4$. (d) Same as (c), but with $\sigma=0.7$. (e) Same as (c), but with $\sigma=0.8$, where the scale of the $y$-axis is enlarged to show that the entire solution diverges.}
    \label{fig:1D_tomo_weight}
\end{figure*}

To verify the accuracy of UQ using FPI-BPINN and fParVI-PINN, we applied the methods to the same synthetic 1D tomography tests as in \cite{Agata2023}, in which a semi-analytical solution is available.
Seismic traveltime tomography is a technique used to determine the seismic velocity structure, i.e., the variation of seismic wave speeds, of the Earth's interior from seismic traveltime observations, i.e., the duration required for a seismic wave to travel from its source to a seismometer.
The basic equation for determining the traveltime is the eikonal equation, which relates the spatial derivative of the traveltime field to the velocity structure as follows:
\begin{eqnarray}
|\nabla T(x,x_\mathrm{s})|^{2} &=& \displaystyle \frac{1}{v^{2}(x)}, \quad \forall\,x \in \Omega \label{eqn:eikonal}\\
T(x_\mathrm{s}, x_\mathrm{s}) &=& 0,\label{eqn:PC}
\end{eqnarray}
where $\Omega$ is a $\mathbb{R}^1$ domain. $T(x,x_\mathrm{s})$ is the traveltime at point $x$ from source $x_\mathrm{s}$, $v(x)$ is the velocity field defined on $\Omega$, and $\nabla$ denotes the gradient operator.
$T$ and $v$ are represented by $f_u$ and $f_m$, respectively.
The second equation defines the point source condition.
We considered the synthetic traveltime data set $\mathcal{D}$ as
\begin{equation}
\mathcal{D} = \{x_i,x_{{\rm s}\,i},T(x_i,x_{{\rm s}\,i})\}_{i=1}^{N_d}
\end{equation}
which was used for the Bayesian estimation of the velocity structure.
We set a simple true velocity model, with a constant velocity of 1\,km/s in the 1D domain defined by $0 \leq x \leq 1.2\,{\rm km}$, to test the UQ performance (see Fig. \ref{fig:1D_tomo_weight} (a), for instance).
We employed such a simple structure because the focus here is the ability of UQ, not the estimation of velocity.
We distributed ten points, serving both as receivers and sources, in two regions defined by the intervals $0.2 \leq x \leq 0.4\,{\rm km}$ and $0.8 \leq x \leq 1\,{\rm km}$, using 0.05\,km spacing.
We referred to the five points in each of the above intervals as Group 1 and Group 2, respectively.
We only considered ray paths of seismic waves between points within the same group.
Therefore, the number of traveltime data points was $5\times4+5\times4=40$.
No ray paths exist in the intervals $0 \leq x \leq 0.2\,{\rm km}$, $0.4 \leq x \leq 0.8\,{\rm km}$, and $1 \leq x \leq 1.2\,{\rm km}$, in which the uncertainty of velocity estimation is expected to be closer to that given by the prior.
The constant velocity readily gives the synthetic traveltime between points analytically.
We did not add artificial noise to the traveltimes, to perform test experiments in an ideal situation.
We set the prior probability in the velocity space as a GP with $\mu_{\mathcal{GP}}(x)=1\,{\rm km/s}$ and $a=0.1\,{\rm km/s}$.
We set two cases of the correlation length scale, $l_\mathcal{GP}=0.075\,{\rm km}$ and $0.15\,{\rm km}$, for comparison.
The likelihood function was given as an IID zero-mean Gaussian noise distribution with standard deviation $\sigma_e=5\times10^{-3}$\,s, and the data errors of the traveltime observations were assumed to be uncorrelated, yielding
\begin{equation}
p(\mathcal{D}| \textbf{m}) = \frac{1}{Z}\exp\left(-\frac{1}{2\sigma_e^2} \sum_{i=1}^{40} \left(f_u(x_i,x_{{\rm s}\,i}) - T(x_i,x_{{\rm s}\,i})\right)^2\right),
\end{equation}
where $i$ is the index of the traveltime data and $Z$ is the normalization constant.

In all the experiments, we used FCNNs with RFF.
We applied the Mish activation function to each layer, except for the output layer, where a linear activation was specified.
In the base case, we used two hidden layers for both $f_u$ representing the traveltime solution and $f_{m}$ representing the velocity structure, with 50 and 30 hidden units per layer, respectively.
For $f_m$, we also considered another setting with three hidden layers and 50 hidden units per layer for comparison.
For FPI-BPINN, we first learned $f_m$ from the functional prior of the Gaussian process.
The result of the learning example of the 1D Gaussian process presented in Section \ref{sct:RFF} was directly applied to the case with $l_\mathcal{GP}=0.15\,{\rm km}$.
We conducted training for $l_\mathcal{GP}=0.075\,{\rm km}$ in the same way.
We initialized $\boldsymbol{\theta}_{m}$ based on the initial ${\boldsymbol \mu}$ and ${\boldsymbol \sigma}$ (see Algorithm \ref{alg:fpi_bpinn}) and initialized $\boldsymbol{\theta}_{u}$ using He's method \cite[]{He2015}.
In the learning process of the prior, we used 140 equally distributed points in the domain $-0.1 \leq x \leq 1.3\,{\rm km}$ as data points.
We prepared 10,000 samples from the GP prior, of which we used 9,000 samples for training and 1,000 samples for validation.
The batch size was 1,000, and the number of epochs $L_{\rm prior}$ was 50.
As discussed in Section \ref{sec:FPI-BPINN}, we used preconditioned SGLD+R (pSGLD+R) for the Bayesian estimation of the weight parameters $\theta_m$.
We sampled using pSGLD+R with 16 particles for $L_\mathrm{post}=$2,000 steps, with $n_\mathrm{burnin}=$1,000 steps considered as the burn-in period, except for the case of the NN with three hidden layers and $l_\mathcal{GP}=0.075\,{\rm km}$, in which $L_\mathrm{post}=$3,000 and $n_\mathrm{burnin}=$2000.
Samples were taken every 100 steps, resulting in 160 samples in total.
We fixed the step size of pSGLD+R to $10^{-3}$.
For fParVI-PINN, we adopted fSVGD with 128 particles.
We used the Adam optimizer \cite[]{Kingma2015} with an initial learning rate of $10^{-2}$ to determine $\epsilon_l$, and the number of iterations $L$ for fParVI-PINN was 1,500, except for the case of the NN with three hidden layers and $l_\mathcal{GP}=0.075\,{\rm km}$, in which the initial learning rate was 5$\times$10$^{-3}$ and $L=2,000$.
We initialized both $\boldsymbol{\theta}_{m}$ and $\boldsymbol{\theta}_{u}$ using He's method \cite[]{He2015}.
As a result, 128 samples were obtained.
Although the experimental setting for fParVI-PINN was basically the same as in \cite{Agata2023}, we adjusted the setting for RFF following Equation \ref{eqn:characteristic_frequency}, as in FPI-BPINN.
In the experiments using both methods, we used full-batch traveltime data.
We took the evaluation points used to define $\mathbf{m}$ using NNs and the collocation points for PINN training for $\boldsymbol{\theta}_{u}$ to be the same, and generated them in each iteration by random sampling in the target domain.
We used 200 evaluation points.
We conducted the PINN training using the L-BFGS algorithm \cite[]{Liu1989} for 10 epochs in each iteration.
See Appendix \ref{sec:appendix_1D_eikonal} for details of the PINN training.

Linearized traveltime tomography estimates velocity perturbation from a reference model using a Taylor series expansion.
When a conjugate pair of the prior and posterior PDF is adopted, such as Gaussian distributions, Bayesian linear regression for linearized tomography provides a semi-analytical solution for the posterior PDF (see \cite{Agata2023}).
We treated this solution as the ground truth of the posterior probability.

We first demonstrate the importance of taking a functional-prior-based approach when aiming for Bayesian estimation consistent with physical intuition, by contrasting it with Bayesian estimation based on the weight space of an NN.
In the semi-analytical solution of the functional-prior-based Bayesian estimation, the uncertainty in regions around Group 1 or Group 2, which are associated with ray paths, is relatively small (Fig. \ref{fig:1D_tomo_weight} (a)(b)).
The uncertainty elsewhere is larger because of the absence of ray paths.
As the spatial correlation increases from $l_\mathcal{GP}=0.075\,{\rm km}$ to $0.15\,{\rm km}$, this uncertainty in the regions without ray paths decreases accordingly.
In this sense, we obtained UQ results that are consistent with physical intuition.
In the weight-space Bayesian estimation using an NN without RFF, we considered the IID normal distribution, which is simple and widely used, as the weight-space prior.
In this case, we set the standard deviation vector of the weight-space prior to ${\boldsymbol \sigma}=\sigma\textbf{1}$, where $\textbf{1}$ is a vector of ones.
We varied the value of $\sigma$ in the Bayesian estimations.
The change of $\sigma=$\,0.4 to 0.7 did not increase the uncertainty in the region between observation Groups 1 and 2 sufficiently (Fig. \ref{fig:1D_tomo_weight} (c)(d)).
The standard deviation in the region in these cases remained less than 0.05\,km/s.
Considering that no ray paths pass through this region, it is reasonable to regard this as an underestimation.
Further increasing $\sigma$ to 0.8 did not selectively enlarge the uncertainty between Groups 1 and 2.
Instead, the entire solution diverged (Fig. \ref{fig:1D_tomo_weight} (e)).
These results demonstrate that it is difficult to obtain UQ results consistent with physical intuition when taking a weight-prior-based approach to Bayesian estimation based on commonly adopted simple prior distributions.

\begin{figure*}[htbp]
    \centering
    \includegraphics[width=0.82\textwidth, bb = 0 0 345 507]{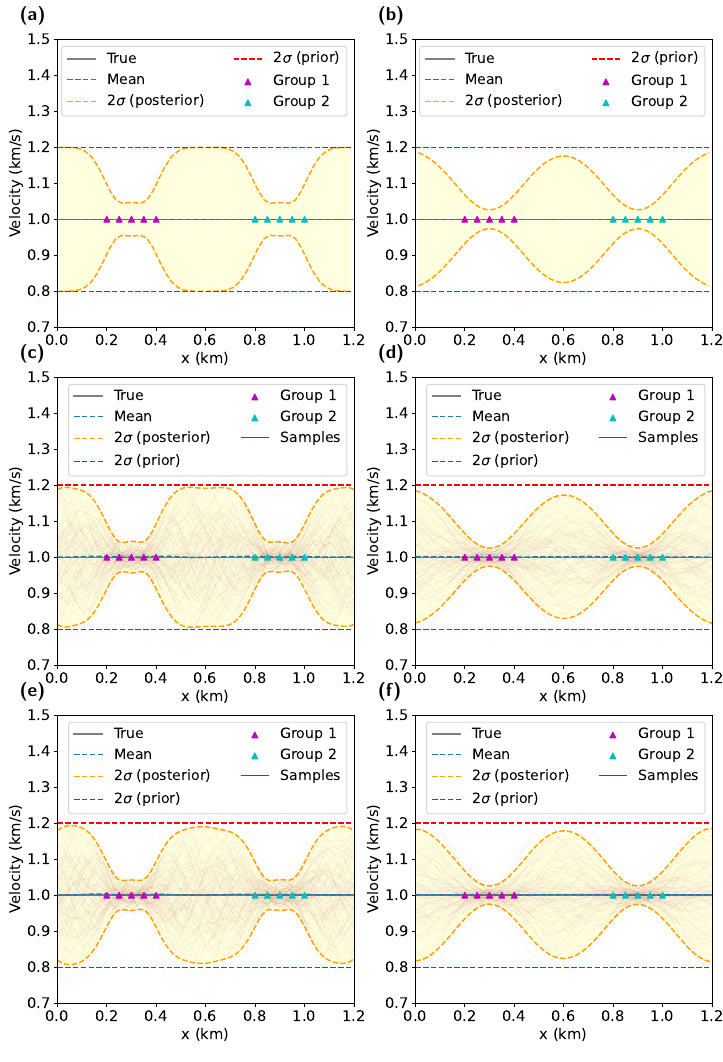}
    \caption{Posterior velocity distributions estimated by fParVI-PINN for the 1D seismic traveltime tomography problem. Each panel shows the posterior mean and standard deviation of the velocity. (a) Semi-analytical posterior for $l_\mathcal{GP}=0.075\,{\rm km}$. (b) Same as (a), but for $l_\mathcal{GP}=0.15\,{\rm km}$. (c) fParVI-PINN result for $l_\mathcal{GP}=0.075\,{\rm km}$. (d) fParVI-PINN result for $l_\mathcal{GP}=0.15\,{\rm km}$. (e) fParVI-PINN result for $l_\mathcal{GP}=0.075\,{\rm km}$ obtained using a larger NN for $f_m$ with three hidden layers and 50 hidden units per layer. (f) Same as (e), but for $l_\mathcal{GP}=0.15\,{\rm km}$.}
    \label{fig:1D_tomo_fParVI}
\end{figure*}

\begin{figure*}[htbp]
    \centering
    \includegraphics[width=0.82\textwidth, bb = 0 0 345 507]{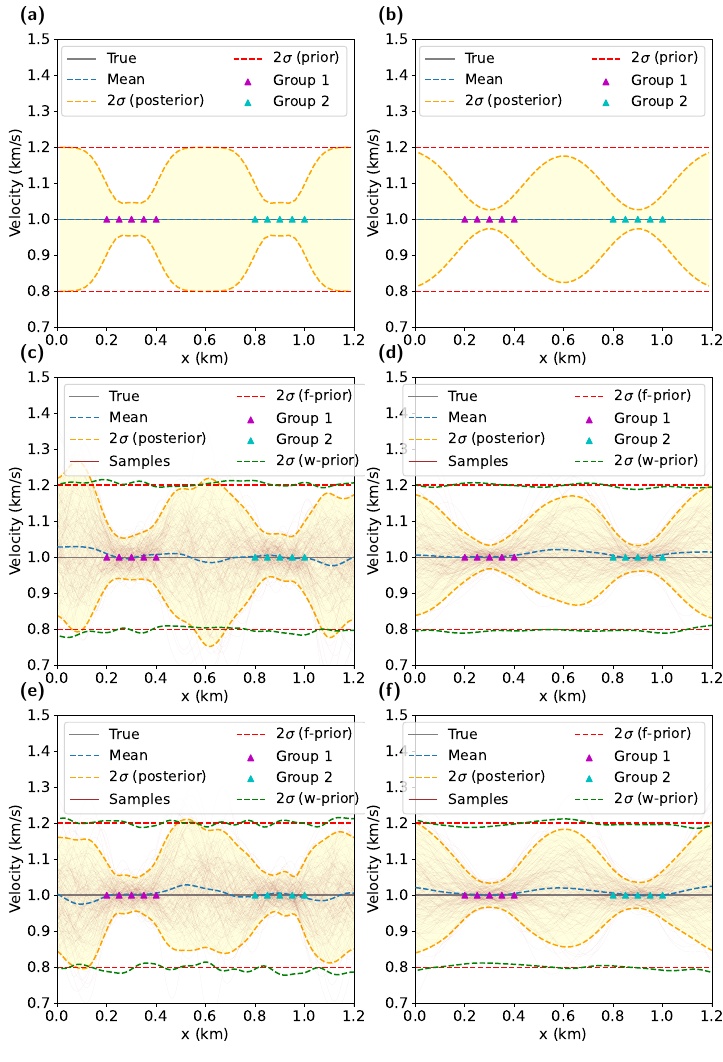}
    \caption{Posterior velocity distributions estimated by FPI-BPINN for the 1D seismic traveltime tomography problem. Each panel shows the posterior mean and standard deviation of the velocity. (a) Semi-analytical posterior for $l_\mathcal{GP}=0.075\,{\rm km}$. (b) Same as (a), but for $l_\mathcal{GP}=0.15\,{\rm km}$. (c) FPI-BPINN result for $l_\mathcal{GP}=0.075\,{\rm km}$. (d) FPI-BPINN result for $l_\mathcal{GP}=0.15\,{\rm km}$. (e) FPI-BPINN result for $l_\mathcal{GP}=0.075\,{\rm km}$ obtained using a larger NN for $f_m$ with three hidden layers and 50 hidden units per layer. (f) Same as (e), but for $l_\mathcal{GP}=0.15\,{\rm km}$.}
    \label{fig:1D_tomo_FPI}
\end{figure*}

Both FPI-BPINN and fParVI-PINN provided accurate results for functional-prior-based Bayesian estimation using neural networks.
In both cases of $l_\mathcal{GP}=0.075\,{\rm km}$ and $0.15\,{\rm km}$, the results from FPI-BPINN and fParVI-PINN agreed well with the semi-analytical solution of the linearized tomography (Fig. \ref{fig:1D_tomo_fParVI} and \ref{fig:1D_tomo_FPI}).
fParVI-PINN provided particularly good results, which can be attributed to the direct incorporation of the functional prior of the Gaussian process (Fig. \ref{fig:1D_tomo_fParVI} (a)-(d)).
Although these results are similar to those presented previously \cite[]{Agata2023}, the accuracy was improved because of the theoretically optimal choice of the characteristic frequency $\tau$ in the RFF.
By contrast, FPI-BPINN with pSGLD+R yielded 160 samples and provided accurate results despite the smaller number of particles (16 particles) than in fParVI-PINN (128 particles and samples) (Fig. \ref{fig:1D_tomo_FPI} (a)-(d)).
This is because the flexibility of FPI-BPINN in the choice of Bayesian estimation method allows for the use of an efficient Bayesian estimation method, such as pSGLD+R.
We also found that our fpBPINN methods are not very sensitive to the structure of the PDE-parameter NN, as demonstrated by the corresponding results obtained using an NN with three hidden layers and 50 hidden units per layer for $f_m$ (Fig. \ref{fig:1D_tomo_fParVI} (e)(f) and \ref{fig:1D_tomo_FPI} (e)(f)).
To evaluate the results more quantitatively, we calculated MMD, the same quantity used for the loss function in the prior training in FPI-BPINN, for each case to compare the accuracy of the results against the corresponding semi-analytical solution.
Because MMD measures the discrepancy between the full estimated and reference posterior distributions, it provides a general assessment of posterior accuracy, including differences in posterior spread and uncertainty structure.
Both FPI-BPINN and fParVI-PINN yielded significantly smaller MMD values than the best-performing case of the weight-space IID prior without RFF, in which $\sigma=0.3$ was chosen by grid search for both $l_\mathcal{GP}=0.075$ and $l_\mathcal{GP}=0.15$ (Table \ref{tab:prior_method_comparison}).
The comparison also shows that fParVI-PINN overall resulted in better performance than FPI-BPINN.

\begin{figure*}[htbp]
    \centering
    \includegraphics[width=0.82\textwidth, bb = 0 16 345 195]{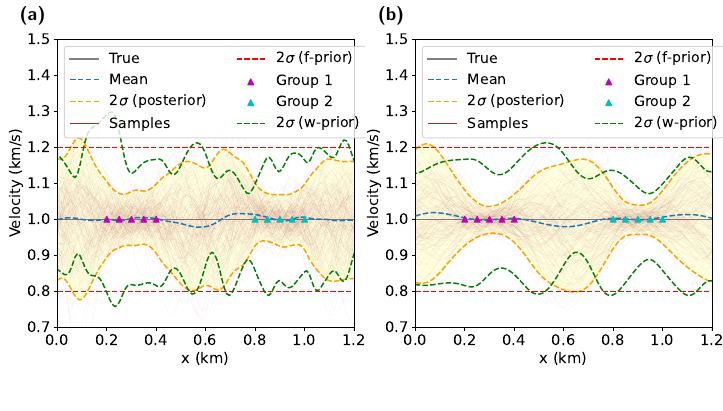}
    \caption{Posterior velocity distributions for 1D seismic traveltime tomography obtained using simple zero-mean IID Gaussian priors in NN weight space with an FCNN with RFF. Each panel shows the posterior mean and standard deviation of the velocity. (a) Result with the characteristic frequency set for $l_\mathcal{GP}=0.075\,{\rm km}$. (b) Result with the characteristic frequency set for $l_\mathcal{GP}=0.15\,{\rm km}$.}
    \label{fig:1D_tomo_RFF}
\end{figure*}

Interestingly, applying the weight-space IID normal distribution prior to an NN with RFF yielded results close to those of functional-prior-based Bayesian estimation.
We considered two cases of the characteristic frequency $\tau$, which were set based on the correlation length $l_\mathcal{GP}=0.075\,{\rm km}$ and $l_\mathcal{GP}=0.15\,{\rm km}$.
These additional cases can be interpreted as partially functional-prior-informed.
Based on the best-performing standard deviation $\sigma=0.7$ chosen for both values of $l_\mathcal{GP}$ by grid search, the UQ results showed a tendency similar to those of the semi-analytical solutions (Fig. \ref{fig:1D_tomo_RFF}).
The MMD value was much smaller than that in the best case without RFF and nearly comparable to that of FPI-BPINN (Table \ref{tab:prior_method_comparison}).
These results suggest that embedding frequency features into the neural network according to those of the functional-prior distributions is crucial for obtaining an accurate approximation of the target posterior PDF.

\subsection{2D Darcy flow}
\label{sec:2D_Darcy}

\begin{figure*}[htbp]
    \centering
    \includegraphics[width=1.0\textwidth]{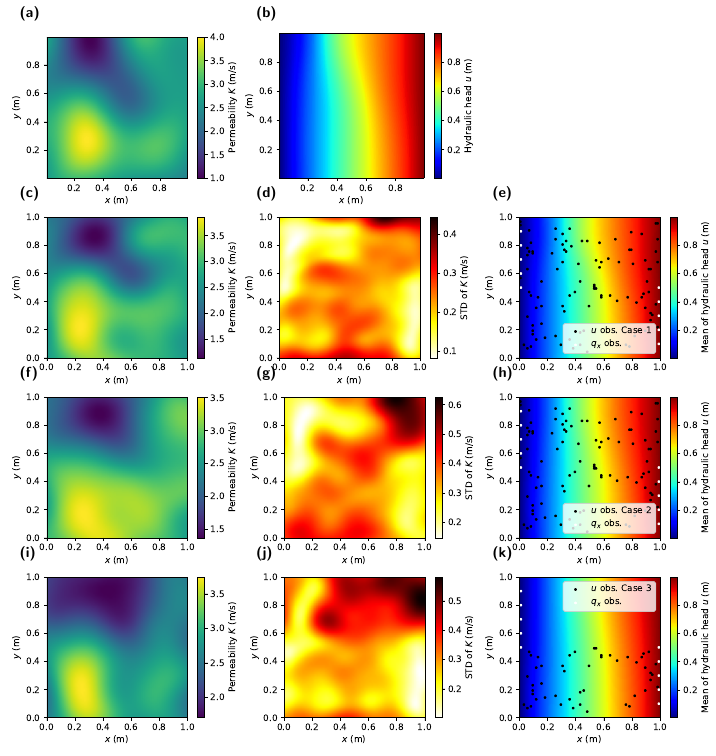}
    \caption{Results of 2D Darcy-flow permeability estimation obtained using FPI-BPINN for three observation patterns. (a) True permeability field $K$. (b) Pressure field $u$ generated from the true permeability field. (c)--(e) Posterior mean of $K$, posterior standard deviation of $K$, and posterior mean of $u$ for Pattern 1, respectively. (f)--(h) Same as (c)--(e), but for Pattern 2. (i)--(k) Same as (c)--(e), but for Pattern 3. The pressure observation points and boundary flux data $q_x$ used in each pattern are also shown in (e), (h), and (k).}
    \label{fig:2D_Darcy_FPI}
\end{figure*}

We considered the estimation of the permeability field in a 2D Darcy flow problem and investigated the applicability of fpBPINN methods to more realistic problems.
The equations for Darcy's law are given as follows:
\begin{eqnarray}
\mathbf{x} = (x, y) \in \Omega = [0,1]^2\\
- \nabla \cdot ( K(\mathbf{x}) \nabla u(\mathbf{x}) ) = 0 , \quad \mathbf{x} \in \Omega \label{eq:darcy_flow_pde}
\end{eqnarray}
where $K$ and $u$ are the permeability field and the pressure field defined on the 2D domain $\Omega$, respectively.
$K$ and $u$ are represented by $f_m$ and $f_u$, respectively.
The boundary conditions are given as follows:
\begin{eqnarray}
u(\mathbf{x}) = 0 , \quad \mathbf{x} \in \Gamma_L := \{ \mathbf{x} \in \partial \Omega | x = 0 \}\\
u(\mathbf{x}) = 1 , \quad \mathbf{x} \in \Gamma_R := \{ \mathbf{x} \in \partial \Omega | x = 1 \}\\
\frac{\partial u}{\partial n}(\mathbf{x}) = 0 , \quad \mathbf{x} \in \Gamma_N := \{ \mathbf{x} \in \partial \Omega | y = 0, 1 \}
\end{eqnarray}
The synthetic pressure and flux data set $\mathcal{D}$ was given as follows:
\begin{eqnarray}
    \mathcal{D} = \{\mathbf{x}_i^u, u(\mathbf{x}_i^u)\}_{i=1}^{N_u} \cup \{\mathbf{x}_j^q, q_x(\mathbf{x}_j^q)\}_{j=1}^{N_q},
\end{eqnarray}
where $\mathbf{x}_i^u$ and $\mathbf{x}_j^q$ are the evaluation points for pressure and flux data, respectively.
We set a true permeability field (Fig. \ref{fig:2D_Darcy_FPI} (a)) and calculated the pressure field and flux using the finite difference method, with the results used as the synthetic data (Fig. \ref{fig:2D_Darcy_FPI} (b)).
No (semi-)analytical solution was available as a reference for this problem.
Instead, we varied the observation patterns and examined whether the resulting changes in the estimated fields were qualitatively reasonable, in order to assess the robustness of the proposed methods with respect to different data configurations.
In Pattern 1, we distributed $N_u=$\,100 points randomly in the domain (Fig. \ref{fig:2D_Darcy_FPI} (e)) with a standard deviation of data error $\eta_{u}=$\,0.01.
In Pattern 2, we used the same $N_u$ points as in Pattern 1 (Fig. \ref{fig:2D_Darcy_FPI} (h)), but with $\eta_{u}=$\,0.05.
In Pattern 3, we distributed $N_u=$\,50 points in the bottom half of the domain (Fig. \ref{fig:2D_Darcy_FPI} (k)) with the same $\eta_{u}=$\,0.01.
To provide scale information for the permeability field, we also used $N_q=$\,10 points for flux data at the $x=0$ and $x=1$ boundaries of the domain (plotted as $q_x$ in Fig. \ref{fig:2D_Darcy_FPI} (e), (h), and (k)) with data error $\eta_{q}=$\,0.05.
We added artificial Gaussian noise according to the data error level for all synthetic data.
We set the prior probability of the permeability field as a 2D GP with $\mu(x)=2.4\,{\rm km/s}$, $\sigma_a=0.8\,{\rm km/s}$, and $l_\mathcal{GP}=0.3\,{\rm km}$.
The likelihood function was given as an IID zero-mean Gaussian noise distribution, written as
\begin{equation}
p(\mathcal{D}| \textbf{m}) = \frac{1}{Z} \exp\left(-\frac{1}{2\eta_{u}^2} \sum_{i=1}^{N_u} (f_u(x_i) - u(x_i))^2 -\frac{1}{2\eta_{q}^2} \sum_{i=1}^{N_q} (f_q(x_i) - q_x(x_i))^2 \right),
\end{equation}
where $i$ is the index of the pressure and flux data.
$f_q$ is an NN-based approximation of the flux, whose details are described in Appendix \ref{sec:appendix_2D_Darcy}.

Similarly to the 1D case, we used FCNNs with RFF with the Mish activation function.
We used NNs with two hidden layers and 50 hidden units per layer for $f_u$ and $f_{m}$, which correspond to the pressure solution and permeability structure, respectively.
For FPI-BPINN, we first learned $f_m$ from the functional prior of the 2D Gaussian process with $l_\mathcal{GP}=0.3\,{\rm km}$.
The initialization strategy was the same as in the 1D case.
In every iteration of the learning process of the prior, we newly sampled 1,024 randomly distributed points in the domain.
1,024 samples from the GP prior for calculating the MMD were also taken at every iteration.
The number of epochs $L_{\rm prior}$ was 2,000.
For FPI-BPINN, we conducted posterior sampling using pSGLD+R with 16 particles for $L_\mathrm{post}=$4,000 steps with $n_\mathrm{burnin}=$2,000.
Samples were taken every 100 steps, resulting in 320 samples in total.
We fixed the step size of pSGLD+R to $3\times10^{-4}$.
For fParVI-PINN, we used fGFSF, the function-space version of Gradient Flow with Smoothed Test Functions (GFSF, \cite[]{Liu2019ICML}), which is a variant of fParVI.
The algorithm of GFSF is similar to that of SVGD, but the gradient term is not smoothed by the kernel as in SVGD.
This makes the computation of the derivative of $K$ included in Equation \ref{eq:darcy_flow_pde} in the fParVI framework easier.
We used the Adam optimizer \cite[]{Kingma2015} with an initial learning rate of $10^{-3}$ to determine $\epsilon_l$, and the number of iterations $L$ for fParVI-PINN was 1,000.
The initialization strategy was the same as in the 1D case.
As a result, 128 samples were obtained.
In the experiments using both methods, we used full-batch pressure and flux data.
We took 2,000 evaluation points in $\Omega$ to define $\mathbf{m}$ using NNs, and used the same points as the collocation points for PINN training for $\boldsymbol{\theta}_{u}$.
We took 200 collocation points for the boundary conditions.
These points were generated in each iteration by random sampling in the domain.
We conducted the PINN training using the SOAP optimizer \cite[]{Vyas2025ICLR}, which has been reported to be effective for improving the convergence of PINNs \cite[]{Wang2025NeurIPS}, for 100 epochs in each iteration.
See Appendix \ref{sec:appendix_2D_Darcy} for details of the PINN training.

\begin{figure*}[htbp]
    \centering
    \includegraphics[width=1.0\textwidth]{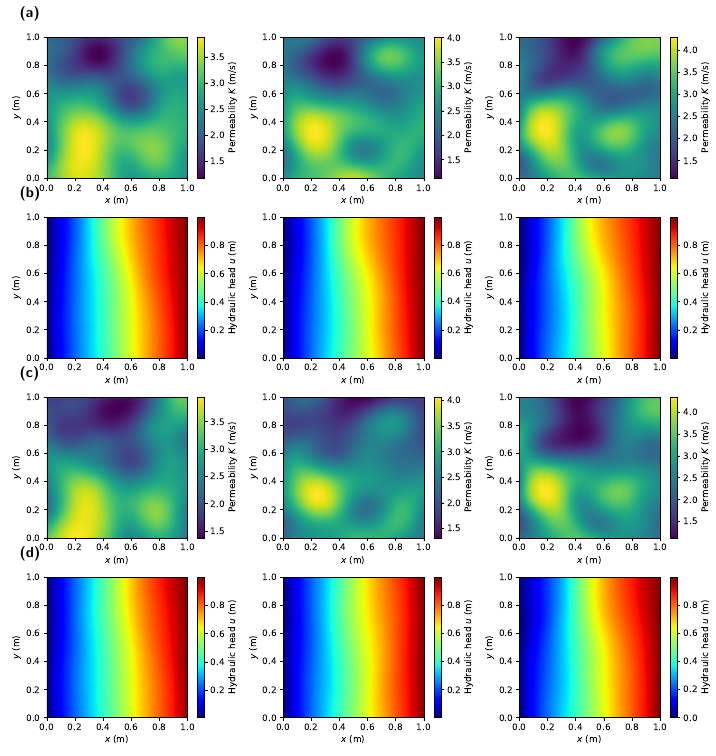}
    \caption{Examples of posterior samples obtained using FPI-BPINN for the 2D Darcy-flow problem. (a) Three samples of the permeability field $K$ for Pattern 1. (b) The corresponding pressure field samples $u$ for Pattern 1. (c) Three samples of $K$ for Pattern 3. (d) The corresponding samples of $u$ for Pattern 3.}
    \label{fig:2D_Darcy_FPI_samples}
\end{figure*}

\begin{figure*}[htbp]
    \centering
    \includegraphics[width=1.0\textwidth]{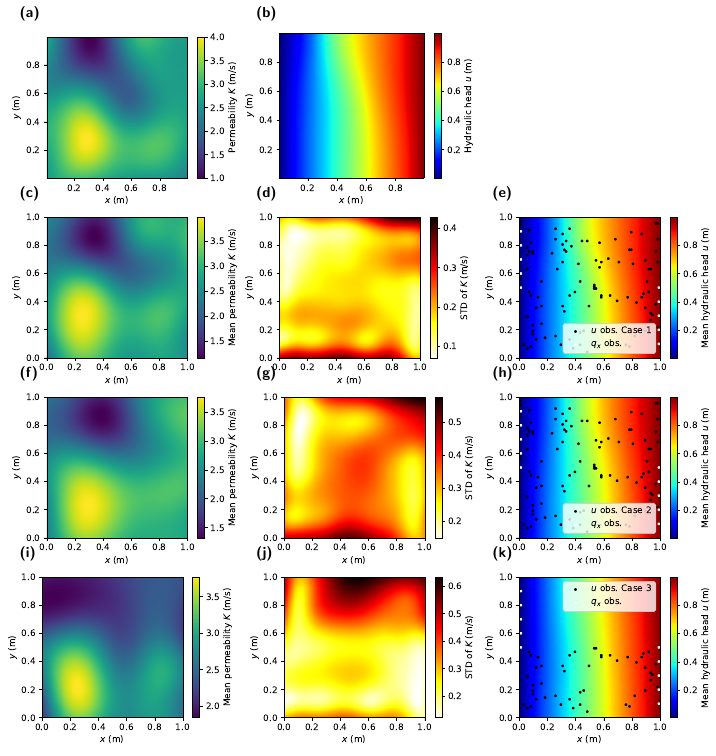}
    \caption{Results of 2D Darcy-flow permeability estimation obtained using fParVI-PINN for three observation patterns. (a) True permeability field $K$. (b) Pressure field $u$ generated from the true permeability field. (c)--(e) Posterior mean of $K$, posterior standard deviation of $K$, and posterior mean of $u$ for Pattern 1, respectively. (f)--(h) Same as (c)--(e), but for Pattern 2. (i)--(k) Same as (c)--(e), but for Pattern 3. The pressure observation points and boundary flux data $q_x$ used in each pattern are also shown in (e), (h), and (k).}
    \label{fig:2D_Darcy_fParVI}
\end{figure*}

\begin{figure*}[htbp]
    \centering
    \includegraphics[width=1.0\textwidth]{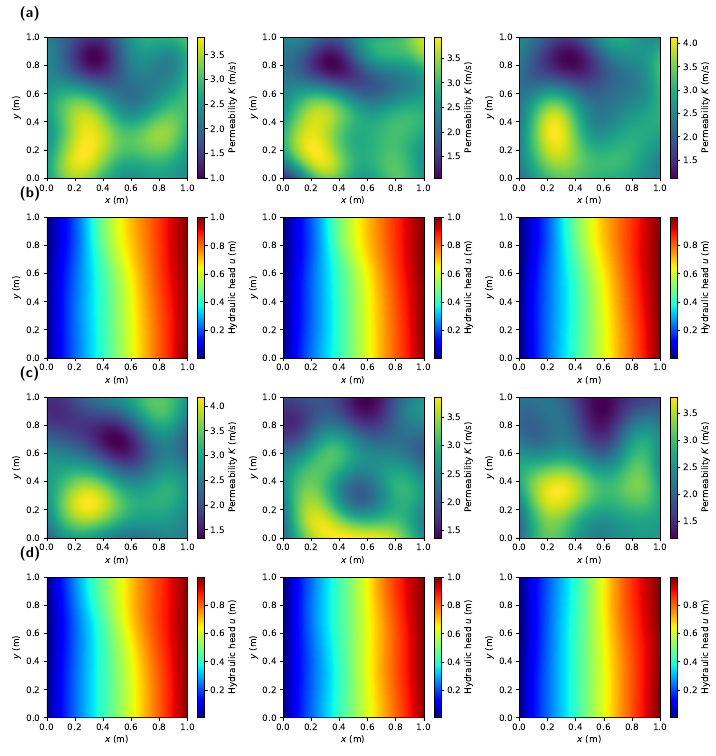}
    \caption{Examples of posterior samples obtained using fParVI-PINN for the 2D Darcy-flow problem. (a) Three samples of the permeability field $K$ for Pattern 1. (b) The corresponding pressure field samples $u$ for Pattern 1. (c) Three samples of $K$ for Pattern 3. (d) The corresponding samples of $u$ for Pattern 3.}
    \label{fig:2D_Darcy_fParVI_samples}
\end{figure*}

Using FPI-BPINN, we obtained samples of $\boldsymbol{\theta}_{m}$ and $\boldsymbol{\theta}_{u}$, each of which outputs $K$ and $u$ at the grid evaluation points in the domain.
The posterior mean of $K$ and the mean of $u$ calculated from these samples corresponding to each $K$ sample agreed well with the true field in Patterns 1 and 2 (Fig. \ref{fig:2D_Darcy_FPI} (c), (e), (f), and (h)).
On the other hand, Pattern 1 showed a significantly smaller standard deviation, which is a metric for uncertainty, than Pattern 2 except for the region near the flux data.
This was a reasonable result reflecting the condition that the difference in the noise level of the pressure data was significant between Patterns 1 and 2, while that of the flux data was the same.
In Pattern 3, which lacks pressure data in the top half of the domain, the posterior mean of $K$ and the mean of $u$ agreed well only with the true field in the bottom half of the domain and in the region near the flux data.
The map of the standard deviation of $K$ showed a significant contrast between large and small values in the top and bottom halves of the domain, respectively.
Fig. \ref{fig:2D_Darcy_FPI_samples} (a)(b) and (c)(d) show examples of $K$ and $u$ samples for Patterns 1 and 3, respectively.
The larger variability of $K$ and $u$ in Pattern 3 than in Pattern 1, in particular in the top half of the domain, is apparent.
The distribution of $u$ does not vary significantly among the samples in Pattern 1, reflecting the data coverage over the whole domain.
In the same way, we obtained samples of $K$ and $u$ using fParVI-PINN and calculated the posterior mean and standard deviation.
The results show a similar tendency to those of FPI-BPINN in all patterns (Fig. \ref{fig:2D_Darcy_fParVI} and Fig. \ref{fig:2D_Darcy_fParVI_samples}).
The rough distribution patterns observed in the standard deviation of $K$ in FPI-BPINN are reduced in fParVI-PINN.
We also observed such a difference in the 1D case, probably because of the different strategy for introducing the functional prior.

Overall, the comparison of the three observation patterns suggests that both FPI-BPINN and fParVI-PINN provided reasonable results for Bayesian estimation of the 2D permeability field under the constraints of the Darcy flow PDE.
It took about 18 hours to complete the calculations in each case of FPI-BPINN using eight NVIDIA A100 GPUs on Earth Simulator 4, made available by the Japan Agency for Marine-Earth Science and Technology (JAMSTEC).
In the case of fParVI-PINN, it took less than seven hours to complete the calculations using the same computational facilities.
Note that the computation codes still need to be optimized for further speedup.

\section{Discussion \& Conclusion}

In this study, we investigated Bayesian PDE-constrained inversion based on physics-informed neural networks from the viewpoint of functional priors. Physically meaningful prior assumptions are more naturally described in function space; therefore, we introduced a unified framework, fpBPINN. Within this framework, we considered two complementary approaches: FPI-BPINN, which learns a weight-space prior consistent with a prescribed functional prior and then performs Bayesian inference in weight space, and fParVI-PINN, which performs particle-based variational inference directly in function space. We also showed that, when Gaussian processes are used as functional priors, the introduction of random Fourier features is important for representing the corresponding frequency characteristics in the PDE-parameter NN.
Through one-dimensional seismic traveltime tomography and two-dimensional Darcy-flow permeability inversion, we demonstrated that both approaches provided accurate posterior estimates under PDE constraints. These findings suggest that functional-prior-based formulations provide a promising direction for uncertainty quantification in PINN-based inverse problems.

FPI-BPINN and fParVI-PINN have complementary strengths and limitations in terms of prior representation, flexibility of posterior inference, and computational scalability. The computational complexity of FPI-BPINN with SGLD+R and that of fParVI-PINN is of the same order in the posterior update stage. In both methods, the dominant cost per iteration is the PDE-constrained gradient evaluation and PINN-based update of PDE solution for each particle.  On the other hand, the posterior update of FPI-BPINN in the presented examples is based on SGLD+R and can be interpreted as a multiple-chain SG-MCMC sampler. Therefore, after burn-in, $M$ samples can be retained from each of the $n$ interacting chains, yielding up to $nM$ posterior samples. The Monte Carlo error of posterior expectations is governed by the effective sample size $N_{\rm eff}$, with the ideal rate $O(N_{\rm eff}^{-1/2})$ and $N_{\rm eff}\le nM$, whereas fParVI-PINN represents the posterior directly by $n$ deterministic particles.
An additional advantage of FPI-BPINN is its flexibility after prior learning. Once the weight prior has been learned from the functional prior, the subsequent inference follows the standard BNN framework. In principle, this allows us to adopt a wide range of existing Bayesian inference methods if the computational cost is affordable, including Hamiltonian Monte Carlo. In addition, because the prior-learning step in the present study is based on a sample-based MMD calculation, it may be possible to extend the framework so that priors are learned directly from discrete datasets. This strategy is analogous to approaches explored in Bayesian inference using generative models \cite[]{Meng2022JCP,Li2025arXiv}. By contrast, as demonstrated by the numerical results for the one-dimensional traveltime tomography problem, fParVI-PINN can achieve accurate inference because the prior distribution in function space is incorporated analytically. 
Moreover, an ongoing study applies fParVI-PINN to a real-world three-dimensional problem \cite[]{Agata2026GJI}, suggesting that this approach has practical potential when sufficient parallel computational resources are available. 

We considered only GPs with the RBF kernel as functional priors in all the experiments in this study. The main reason is that a GP with an RBF kernel is highly physically interpretable and the most standard setup in GP regression. Nevertheless, the proposed framework is not restricted to such GPs. For example, application of Mat\'ern kernels is also possible theoretically and investigating the practical effectiveness of introducing them is an interesting direction for future work. In addition, FPI-BPINN can be applied to any stochastic process from which samples can be generated, whereas fParVI-PINN can be applied when the prior can be described in closed form in function space.
The selection of hyperparameters in the functional prior, such as the correlation length and amplitude of the GP, was not investigated systematically in this study. This issue is important because such hyperparameters directly determine prior assumptions on smoothness, correlation scale, and variability of the unknown parameter field. A possible strategy is an empirical-Bayesian approach. For example, in applications of fParVI-PINN to real-world geophysical problems \cite[]{Agata2025SR,Agata2026GJI}, the hyperparameters were determined using the widely applicable Bayesian information criterion (WBIC) \cite[]{Watanabe2013}.

In the present study, we focused on the case in which only a single PDE parameter is unknown. In PINN-based inverse problems, however, multiple PDE parameters are often estimated simultaneously, and the corresponding NN has multiple outputs (e.g., \cite[]{Yin2021CMAME,Kamali2023AB,Fukushima2025JGR-SE}). Extending the present framework to such settings appears to be conceptually straightforward, but its practical applicability to real problems remains to be examined.
In particular, when multiple PDE parameters are inferred simultaneously, an important issue is how to cope with the implicit correlation between the parameters imposed by sharing the same NN weight parameters.

\section*{Acknowledgments}
This study was supported by JSPS KAKENHI Grant Number 25K01084 and ERI JURP 2025-B-01 in Earthquake Research Institute, the University of Tokyo. The calculations were carried out using Earth Simulator at JAMSTEC.

\appendix
\section{Appendix}

\subsection{Derivation of the gradient of the loss function using the adjoint method}
\label{sec:appendix_adjoint}

For ease of subsequent derivations, we present the discrete form of the Lagrangian as
\begin{equation}
\mathcal{L}_d(\boldsymbol{\theta}_u,\boldsymbol{\theta}_m,\boldsymbol{\lambda}_{\mathcal{F}},\boldsymbol{\lambda}_{\mathcal{B}})
=
\mathcal{J}(\boldsymbol{\theta}_u,\boldsymbol{\theta}_m)
+
\sum_{i=1}^{N_{\mathcal{F}}} w_i^{\mathcal{F}} \lambda_{\mathcal{F},i}\,\mathcal{F}_i(\boldsymbol{\theta}_u,\boldsymbol{\theta}_m)
+
\sum_{j=1}^{N_{\mathcal{B}}} w_j^{\mathcal{B}} \lambda_{\mathcal{B},j}\,\mathcal{B}_j(\boldsymbol{\theta}_u),
\label{eq:discrete_lagrangian}
\end{equation}
where $N_{\mathcal{F}}$ and $N_{\mathcal{B}}$ are the numbers of interior and boundary collocation points, respectively, and $w_i^{\mathcal{F}}$ and $w_j^{\mathcal{B}}$ are quadrature weights.
The quantities $\mathcal{F}_i$ and $\mathcal{B}_j$ denote the PDE residual and boundary residual evaluated at the corresponding collocation points.
To derive the gradient with respect to $\boldsymbol{\theta}_m$, we consider the total derivative of $\mathcal{L}_d$:
\begin{equation}
\frac{d \mathcal{L}_d}{d \boldsymbol{\theta}_m}
=
\frac{\partial \mathcal{L}_d}{\partial \boldsymbol{\theta}_m}
+
\frac{\partial \mathcal{L}_d}{\partial \boldsymbol{\theta}_u}
\frac{d \boldsymbol{\theta}_u}{d \boldsymbol{\theta}_m}
+
\frac{\partial \mathcal{L}_d}{\partial \boldsymbol{\lambda}}
\frac{d \boldsymbol{\lambda}}{d \boldsymbol{\theta}_m},
\label{eq:total_derivative_discrete_lagrangian}
\end{equation}
where $\boldsymbol{\lambda}=\{\boldsymbol{\lambda}_{\mathcal{F}},\boldsymbol{\lambda}_{\mathcal{B}}\}$.
Since the forward solution approximately satisfies the PDE and boundary constraints at the collocation points, the term $\partial \mathcal{L}_d / \partial \boldsymbol{\lambda}$ vanishes.
The second term, however, contains the sensitivity $d\boldsymbol{\theta}_u/d\boldsymbol{\theta}_m$, whose explicit evaluation is computationally expensive when the dimension of $\boldsymbol{\theta}_m$ is large.
The key idea of the adjoint method is to choose $\boldsymbol{\lambda}$ such that the dependence on $d\boldsymbol{\theta}_u/d\boldsymbol{\theta}_m$ is eliminated.
This is achieved by satisfying the following condition:
\begin{equation}
\frac{\partial \mathcal{L}_d}{\partial \boldsymbol{\theta}_u} =
\frac{\partial \mathcal{J}}{\partial \boldsymbol{\theta}_u}
+
\sum_{i=1}^{N_{\mathcal{F}}}
w_i^{\mathcal{F}} \lambda_{\mathcal{F},i}
\frac{\partial \mathcal{F}_i}{\partial \boldsymbol{\theta}_u}
+
\sum_{j=1}^{N_{\mathcal{B}}}
w_j^{\mathcal{B}} \lambda_{\mathcal{B},j}
\frac{\partial \mathcal{B}_j}{\partial \boldsymbol{\theta}_u}
:= \boldsymbol{0}.
\label{eq:expanded_discrete_adjoint}
\end{equation}
Once $\boldsymbol{\lambda}$ is obtained by solving this equation, the total derivative reduces to
\begin{equation}
\nabla_{\boldsymbol{\theta}_m} \mathcal{J}
=
\frac{d \mathcal{L}_d}{d \boldsymbol{\theta}_m}
=
\frac{\partial \mathcal{L}_d}{\partial \boldsymbol{\theta}_m}
=
\frac{\partial \mathcal{J}}{\partial \boldsymbol{\theta}_m}
+
\sum_{i=1}^{N_{\mathcal{F}}} w_i^{\mathcal{F}} \lambda_{\mathcal{F},i}
\frac{\partial \mathcal{F}_i}{\partial \boldsymbol{\theta}_m}
+
\sum_{j=1}^{N_{\mathcal{B}}} w_j^{\mathcal{B}} \lambda_{\mathcal{B},j}
\frac{\partial \mathcal{B}_j}{\partial \boldsymbol{\theta}_m}.
\label{eq:gradient_of_loss_function}
\end{equation}
We use the CGNR method to solve the adjoint equation (Equation \ref{eq:expanded_discrete_adjoint}) and obtain $\boldsymbol{\lambda}$.

\subsection{Preconditioning technique for SGLD+R}
\label{sec:preconditioning_sgldr}

SGLD+R, proposed by Gallego \& Insua \cite{Gallego2018NeurIPS}, can be understood as an interacting-particle extension of ordinary SGLD.
Whereas standard multiple-chain SGLD evolves each chain (particle) independently, SGLD+R introduces kernel-based interactions among particles so that they repel each other and do not collapse to the same mode.
As a result, the method improves the exploration of the posterior while retaining the stochastic noise term required for SGLD sampling.
From the viewpoint of SVGD, SGLD+R can be interpreted as SVGD augmented with a noise term so that the resulting interacting-particle dynamics constitutes a valid SG-MCMC sampler.
Following \cite{Gallego2018NeurIPS}, let
\[
\boldsymbol{\Theta}^{k}
:=
\left[
(\boldsymbol{\theta}_{m,1}^{k})^{\top},
\ldots,
(\boldsymbol{\theta}_{m,n_p}^{k})^{\top}
\right]^{\top}
\]
be the concatenated vector of all particles, where $n_p$ is the number of particles and $k$ is the iteration index.
Then, the general interacting SGLD update is written as
\begin{equation}
\boldsymbol{\Theta}^{k+1}
=
\boldsymbol{\Theta}^{k}
-\epsilon_k
\left[
(\mathbf{D}_{k}+\mathbf{Q}_{k})\nabla \mathcal{H}(\boldsymbol{\Theta}^{k})
+
\boldsymbol{\Gamma}_{k}
\right]
+
\boldsymbol{\eta}_{k},
\qquad
\boldsymbol{\eta}_{k}\sim
\mathcal{N}(\mathbf{0},2\epsilon_k \mathbf{D}_{k}),
\label{eq:sgldr_general_appendix}
\end{equation}
where $\mathcal{H}$ denotes the total negative log-posterior energy of the particle system.
Here, $\mathbf{D}_{k}$ is the diffusion matrix induced by the particle interaction kernel, $\mathbf{Q}_{k}$ is a skew-symmetric curl matrix that appears in more general Hamiltonian-type variants, and $\boldsymbol{\Gamma}_{k}$ is the correction term required in the SGLD framework so that the target distribution is preserved.
In SGLD+R, the repulsive force is encoded through the kernel-dependent structure of $\mathbf{D}_{k}$ and $\boldsymbol{\Gamma}_{k}$.
When the noise term is omitted, the update is equivalent to that of SVGD.
For the detailed definitions of $\mathbf{D}_{k}$, $\mathbf{Q}_{k}$, and $\boldsymbol{\Gamma}_{k}$, we refer the reader to \cite[]{Gallego2018NeurIPS}.

In our implementation, we use the SGLD+R case of this framework and set $\mathbf{Q}_{k}=\mathbf{0}$.
To improve convergence, we further introduce a diagonal preconditioner in the same spirit as preconditioned SGLD.
For each particle, we compute the gradient of the negative log-posterior,
\begin{equation}
\mathbf{g}_{i}^{k}
:=
\nabla_{\boldsymbol{\theta}_{m,i}^{k}} \mathcal{J},
\label{eq:psgldr_gradient}
\end{equation}
and form the moving average of the particle-averaged squared gradients as
\begin{equation}
\mathbf{v}^{k}
:=
\beta \mathbf{v}^{k-1}
+
(1-\beta)\frac{1}{n_p}\sum_{i=1}^{n_p}
\left(\mathbf{g}_{i}^{k}\odot \mathbf{g}_{i}^{k}\right),
\label{eq:psgldr_second_moment}
\end{equation}
where $\beta \in [0,1)$ is the decay rate and $\odot$ denotes the elementwise product.
Using this quantity, we define a parameter-wise diagonal preconditioner
\begin{equation}
\mathbf{G}_{k}
:=
{\rm diag}
\left(
\left(\sqrt{\mathbf{v}^{k}}+\lambda \mathbf{1}\right)^{-1}
\right),
\label{eq:psgldr_preconditioner}
\end{equation}
where $\lambda>0$ is a small constant for numerical stability.
This is analogous to RMSprop: parameters with persistently large gradients are updated more conservatively, while flatter directions are assigned larger effective step sizes.

To apply this preconditioner to the whole interacting particle system, we define
\begin{equation}
\bar{\mathbf{G}}_{k}
:=
\mathbf{I}_{n_p}\otimes \mathbf{G}_{k},
\label{eq:psgldr_lifted_preconditioner}
\end{equation}
and replace the diffusion matrix in Equation \ref{eq:sgldr_general_appendix} by
\begin{equation}
\tilde{\mathbf{D}}_{k}
:=
\bar{\mathbf{G}}_{k}\mathbf{D}_{k}.
\label{eq:psgldr_modified_diffusion}
\end{equation}
Correspondingly, the preconditioned SGLD+R update is written as
\begin{equation}
\boldsymbol{\Theta}^{k+1}
=
\boldsymbol{\Theta}^{k}
-\epsilon_k
\left[
\tilde{\mathbf{D}}_{k}\nabla \mathcal{H}(\boldsymbol{\Theta}^{k})
+
\tilde{\boldsymbol{\Gamma}}_{k}
\right]
+
\tilde{\boldsymbol{\eta}}_{k},
\qquad
\tilde{\boldsymbol{\eta}}_{k}\sim
\mathcal{N}(\mathbf{0},2\epsilon_k \tilde{\mathbf{D}}_{k}),
\label{eq:psgldr_update}
\end{equation}
where $\tilde{\boldsymbol{\Gamma}}_{k}$ is the corresponding correction term associated with $\tilde{\mathbf{D}}_{k}$ in the same sense as $\boldsymbol{\Gamma}_{k}$ is associated with $\mathbf{D}_{k}$.
The adaptive update of $\mathbf{G}_{k}$ is used only during the burn-in period.
After burn-in, the fixed preconditioner, denoted by $\tilde{\mathbf{D}}_{k}^{\ast}$, is used in Equation \ref{eq:psgldr_update} instead of $\tilde{\mathbf{D}}_{k}$.

\subsection{Derivation of the characteristic frequency of the random Fourier features}
\label{sec:appendix_characteristic_frequency}

To determine the characteristic frequency of the random Fourier features (RFF), we first confirm the definition used by \cite{Tancik2020}. In Section 3.2 of their paper, the RFF mapping is defined as
\begin{equation}
\gamma(\mathbf{v})
:=
\left[
\cos(2\pi \mathbf{B}\mathbf{v}),
\sin(2\pi \mathbf{B}\mathbf{v})
\right]^T,
\end{equation}
where each element of $\mathbf{B}$ is sampled from $\mathcal{N}(0,\tau^2)$. We derive a theoretical guideline for choosing $\tau$ from the length scale of the target Gaussian process. Suppose that the desired Gaussian process is defined by the RBF kernel
\begin{equation}
k(x)
:=
\exp\left(-\frac{x^2}{l^2}\right),
\end{equation}
where $l$ is the length scale controlling smoothness. By Bochner's theorem, the sampling distribution of the RFF should match the power spectral density, namely the Fourier transform of the kernel. Using the Fourier-transform convention
\begin{equation}
\mathcal{F}\{g(t)\}(f)
:=
\int g(t)e^{-i2\pi f t}\,dt,
\end{equation}
the Fourier transform of the RBF kernel \cite[]{Rasmussen2006} is
\begin{equation}
\mathcal{F}
\left\{
\exp\left(-\frac{x^2}{l^2}\right)
\right\}(f)
=
\sqrt{\pi}l\exp\left(-\pi^2 l^2 f^2\right).
\end{equation}
Ignoring the constant factor with respect to $f$, this expression is proportional to the probability density of a Gaussian distribution with zero mean and standard deviation $\tau_f$,
\begin{equation}
\exp\left(-\frac{f^2}{2\tau_f^2}\right).
\end{equation}
By comparing the exponents, we obtain
\begin{equation}
\frac{f^2}{2\tau_f^2}
=
\pi^2 l^2 f^2,
\end{equation}
which gives
\begin{equation}
\tau_f^2
=
\frac{1}{2\pi^2 l^2}.
\end{equation}
Therefore, the bandwidth of the RFF is determined as
\begin{equation}
\tau
:=
\tau_f
:=
\frac{1}{\sqrt{2}\pi l}.
\end{equation}
This relation shows that the bandwidth parameter of the RFF layer should be chosen as the standard deviation of the frequency distribution corresponding to the target Gaussian-process length scale.

\subsection{PINN training algorithms}
\label{sec:appendix_pinn_training}

\subsubsection{1D eikonal equation}
\label{sec:appendix_1D_eikonal}

In the 1D problem considered here, the spatial coordinate is the scalar $x \in \Omega \subset \mathbb{R}$.
The eikonal equation (Equation~\ref{eqn:eikonal}) corresponds to the PDE constraint $\mathcal{F}(u,m)=0$ (Equation~\ref{eq:pde_constraint}), with $u=T$ (traveltime) and $m=v$ (seismic velocity), and the point source condition (Equation~\ref{eqn:PC}) serves as the boundary condition.
To avoid singularities in the point source condition, we introduce the following factored form following \cite{Smith2021,Waheed2021PINNeik}:
\begin{eqnarray}
T(x,x_s)=T_{0}(x,x_s)\,\tau(x,x_s),
\end{eqnarray}
where $T_{0}(x,x_s)$ is defined as
\begin{eqnarray}
T_{0}(x,x_s)=\left|x-x_s\right|.
\end{eqnarray}
This factorization automatically satisfies the point source condition, so that the boundary loss $\mathcal{L}_{\rm boundary}$ in Equation~\ref{eq:pinn_training} is not required.
In 1D, the eikonal equation residual simplifies to
\begin{eqnarray}
r_{\rm EE}&=&v(x)-\frac{1}{\left|\partial T/\partial x\,(x,x_s)\right|},\label{eqn:residual}
\end{eqnarray}
which corresponds to $\mathcal{F}(u,m)$ in Equation~\ref{eq:pde_constraint}.

The traveltime NN $f_u$ and the velocity NN $f_m$ are specified as
\begin{eqnarray}
T(x,x_s)
&\simeq& f_{u}(x,x_s,\boldsymbol{\theta}_{u})\nonumber\\
&=& T_{0}(x,x_s)/f_{\tau^{-1}}(x,x_s,\boldsymbol{\theta}_{u}),\label{eqn:T}\\
v(x)
&\simeq& f_{m}(x,\boldsymbol{\theta}_{m})\nonumber\\
&=& v_0(x) + f_{v_{\rm ptb}}(x,\boldsymbol{\theta}_{m}),
\end{eqnarray}
where $f_{\tau^{-1}}$ is an NN-based function approximating $1/\tau(x,x_s)$, $v_0(x)$ is the reference velocity set by the user, and $f_{v_{\rm ptb}}(x,\boldsymbol{\theta}_{m})$ approximates the velocity perturbation.
We approximated $1/\tau$ and $v_{\rm ptb}$ using NNs, instead of directly computing $\tau$ and $v$ as done in previous studies \cite[]{Smith2021,Waheed2021PINNeik}, to improve convergence performance.
We used fully connected feed-forward networks to implement both $f_{\tau^{-1}}$ and $f_{v_{\rm ptb}}$.
Further, the reciprocity condition (i.e., $T(x,x_s)=T(x_s,x)$) was imposed following \cite{Grubas2023} by using $\frac{1}{2}\left(f_{\tau^{-1}}(x,x_s,\boldsymbol{\theta}_{u})+f_{\tau^{-1}}(x_s,x,\boldsymbol{\theta}_{u})\right)$ instead of $f_{\tau^{-1}}(x,x_s,\boldsymbol{\theta}_{u})$ in Equation~\ref{eqn:T} to improve the convergence of the eikonal equation solution.

In the PINN training step (Equation~\ref{eq:pinn_training}), with $\boldsymbol{\theta}_m$ fixed, the traveltime NN is trained by minimizing
\begin{eqnarray}
\mathcal{L}_{\rm PDE}
&=& \alpha \sum_{i=1}^{N_c}\left( f_{m}(x_c^{(i)};\boldsymbol{\theta}_{m})-\frac{1}{\left|\partial f_{u}/\partial x\,(x_c^{(i)},x_s^{(i)};\boldsymbol{\theta}_{u})\right|} \right)^{2},\label{eqn:LPDE}
\end{eqnarray}
where $N_c$ is the number of collocation points, $x_c$ denotes their coordinates, and $\alpha$ is a loss weight typically set to $1/N_{c}$.
The collocation points are selected randomly within the 1D target domain $\Omega$ and serve as the evaluation points for the PDE residuals \cite[]{Raissi2019}.

\subsubsection{2D Darcy's law}
\label{sec:appendix_2D_Darcy}

In the 2D problem, the spatial coordinate is $\mathbf{x} = (x, y) \in \Omega = [0,1]^2$.
The Darcy flow equation (Equation~\ref{eq:darcy_flow_pde}) corresponds to the PDE constraint $\mathcal{F}(u, m) = 0$ (Equation~\ref{eq:pde_constraint}), with $u$ denoting the pressure field and $m = K$ denoting the permeability field.
The PDE residual is given by
\begin{eqnarray}
r_{\rm Darcy}
&=&
-\nabla \cdot \bigl( K(\mathbf{x})\,\nabla u(\mathbf{x}) \bigr) \nonumber \\
&=&
-\frac{\partial K}{\partial x}\frac{\partial u}{\partial x}
-\frac{\partial K}{\partial y}\frac{\partial u}{\partial y}
-K\!\left(\frac{\partial^{2} u}{\partial x^{2}}+\frac{\partial^{2} u}{\partial y^{2}}\right),
\label{eqn:darcy_residual}
\end{eqnarray}
which corresponds to $\mathcal{F}(u, m)$ in Equation~\ref{eq:pde_constraint}.
All partial derivatives are evaluated by automatic differentiation.
The pressure NN $f_u$ and the permeability NN $f_m$ are specified as
\begin{eqnarray}
u(\mathbf{x})
&\simeq& f_{u}(\mathbf{x};\boldsymbol{\theta}_{u}), \\
K(\mathbf{x})
&\simeq& f_{m}(\mathbf{x};\boldsymbol{\theta}_{m}),
\end{eqnarray}
where both $f_u$ and $f_m$ are implemented as FCNNs with RFF and the Mish activation function, each having two hidden layers with 50 units per layer.

In the PINN training step (Equation~\ref{eq:pinn_training}), with $\boldsymbol{\theta}_m$ fixed, the pressure NN is trained by minimizing $\mathcal{L}_{\rm PDE} + \mathcal{L}_{\rm boundary}$.
The PDE loss over $N_c$ interior collocation points $\{\mathbf{x}_c^{(i)}\}$ is
\begin{eqnarray}
\mathcal{L}_{\rm PDE}
=
\frac{1}{N_c}\sum_{i=1}^{N_c}
\left(
r_{\rm Darcy}\!\left(\mathbf{x}_c^{(i)};\boldsymbol{\theta}_{u},\boldsymbol{\theta}_{m}\right)
\right)^{2}.
\label{eqn:darcy_LPDE}
\end{eqnarray}
The boundary loss enforces the Dirichlet and Neumann boundary conditions using $N_b$ boundary collocation points $\{\mathbf{x}_b^{(j)}\}$:
\begin{eqnarray}
\mathcal{L}_{\rm boundary}
&=&
\frac{1}{N_b^{D}}\sum_{j \in \Gamma_L \cup \Gamma_R}
\Bigl( f_{u}\!\left(\mathbf{x}_b^{(j)};\boldsymbol{\theta}_{u}\right) - g_D\!\left(\mathbf{x}_b^{(j)}\right) \Bigr)^{2}
\nonumber\\
&&
+\;
\frac{1}{N_b^{N}}\sum_{j \in \Gamma_N}
\left( \frac{\partial f_{u}}{\partial n}\!\left(\mathbf{x}_b^{(j)};\boldsymbol{\theta}_{u}\right) \right)^{2},
\label{eqn:darcy_Lboundary}
\end{eqnarray}
where $g_D(\mathbf{x}) = 0$ on $\Gamma_L$ and $g_D(\mathbf{x}) = 1$ on $\Gamma_R$,
$N_b^D$ and $N_b^N$ are the numbers of Dirichlet and Neumann boundary collocation points, respectively,
and $\partial/\partial n$ denotes the outward normal derivative.
The flux used as supplementary observational data (Section~\ref{sec:2D_Darcy}) is computed as
\begin{eqnarray}
f_{q}\!\left(\mathbf{x};\boldsymbol{\theta}_{u},\boldsymbol{\theta}_{m}\right)
=
-f_{m}\!\left(\mathbf{x};\boldsymbol{\theta}_{m}\right)\,
\frac{\partial f_{u}}{\partial x}\!\left(\mathbf{x};\boldsymbol{\theta}_{u}\right),
\end{eqnarray}
and enters the negative log-posterior $\mathcal{J}$ through the likelihood function.

\clearpage

\begin{algorithm}[htbp]
    \caption{FPI-BPINN with SGLD+R}
    \label{alg:fpi_bpinn}
    \begin{algorithmic}[1]
        \REQUIRE target stochastic process $\mathcal{SP}(\psi)$, negative log-posterior $\mathcal{J}$, PDE losses $\mathcal{L}_{\rm PDE}$ and $\mathcal{L}_{\rm boundary}$, numbers of prior-learning and posterior-inference iterations $L_{\rm prior}$ and $L_{\rm post}$, number of particles $n_p$
        \ENSURE posterior particles $\{\boldsymbol{\theta}_{m,i}^{L_{\rm post}}\}_{i=1}^{n_p}$ approximating $p(\boldsymbol{\theta}_m | \mathcal{D})$
        \STATE {\bf Stage I: Learning the prior distribution of ${\boldsymbol{\theta}_m}$ by MMD minimization}
        \STATE Initialize $\boldsymbol{\mu}=\mathbf{0}$; set $\boldsymbol{\sigma}$ from He's variance for weights and from one-tenth of that scale for biases
        \FOR{$l=0,\ldots,L_{\rm prior}-1$}
            \STATE Draw $\{\boldsymbol{\theta}_{m}^{(i)}\}_{i=1}^{N} \sim \mathcal{N}(\boldsymbol{\mu}, {\rm diag}(\boldsymbol{\sigma}^{2}))$
            \STATE Compute BNN samples $\{f_m^{(i)}\}_{i=1}^{N}$ by forward propagation
            \STATE Draw function-space samples $\{m_{\mathcal{SP}}^{(i)}\}_{i=1}^{N} \sim \mathcal{SP}(\psi)$
            \STATE Evaluate $\mathcal{L}_{\rm MMD}$ using Equation \ref{eq:mmd_loss}
            \STATE Update $\boldsymbol{\mu}$ and $\boldsymbol{\sigma}$ by stochastic gradient descent
        \ENDFOR
        \STATE Set learned prior parameters $\boldsymbol{\mu}^{\ast} \leftarrow \boldsymbol{\mu}$ and $\boldsymbol{\sigma}^{\ast} \leftarrow \boldsymbol{\sigma}$
        \STATE {\bf Stage II: Posterior inference by particle-based SGLD+R}
        \STATE Draw initial particles $\{\boldsymbol{\theta}_{m,i}^{0}\}_{i=1}^{n_p} \sim \mathcal{N}(\boldsymbol{\mu}^{\ast}, {\rm diag}((\boldsymbol{\sigma}^{\ast})^{2}))$
        \FOR{$l=0,\ldots,L_{\rm post}-1$}
            \FOR{$i=1,\ldots,n_p$}
                \STATE Obtain $\boldsymbol{\theta}_{u,i}^{l}$ by PINN training with fixed $\boldsymbol{\theta}_{m,i}^{l}$:
                \STATE \hspace{1em}$\displaystyle \boldsymbol{\theta}_{u,i}^{l} = \argmin_{\boldsymbol{\theta}_u} \mathcal{L}_{\rm PDE}(\boldsymbol{\theta}_u, \boldsymbol{\theta}_{m,i}^{l}) + \mathcal{L}_{\rm boundary}(\boldsymbol{\theta}_u)$
                \STATE Solve the adjoint equation in Section 2.2 (Equation \ref{eq:expanded_discrete_adjoint})
                \STATE Compute $\mathbf{g}_{i}^{l} = \nabla_{\boldsymbol{\theta}_m}\mathcal{J}(\boldsymbol{\theta}_{u,i}^{l}, \boldsymbol{\theta}_{m,i}^{l})$ using Equation \ref{eq:gradient_of_loss_function}
            \ENDFOR
            \STATE Update $\{\boldsymbol{\theta}_{m,i}^{l}\}_{i=1}^{n_p}$ by one SGLD+R step using $\{\mathbf{g}_{i}^{l}\}_{i=1}^{n_p}$
        \ENDFOR
    \end{algorithmic}
\end{algorithm}

\begin{algorithm}[htbp]
    \caption{fParVI-PINN}
    \label{alg:fparvi_pinn}
    \begin{algorithmic}[1]
        \REQUIRE functional prior $p(\mathbf{m})$, particle number $n_p$, evaluation points ${\bf X}_v$, posterior-inference iterations $L$
        \ENSURE particle approximation of the posterior distribution in function space and weight space, $\{(\mathbf{m}_{i}^{L},\boldsymbol{\theta}_{m,i}^{L})\}_{i=1}^{n_p}$
        \STATE Initialize $\{\boldsymbol{\theta}_{m,i}^{0}\}_{i=1}^{n_p}$
        \FOR{$l=0,\ldots,L-1$}
            \FOR{$i=1,\ldots,n_p$}
                \STATE Evaluate the current function-space particle $\mathbf{m}_{i}^{l}=f_m({\bf X}_v;\boldsymbol{\theta}_{m,i}^{l})$
                \STATE Obtain $\boldsymbol{\theta}_{u,i}^{l}$ by PINN training with fixed $\mathbf{m}_{i}^{l}$ (equivalently, fixed $\boldsymbol{\theta}_{m,i}^{l}$):
                \STATE \hspace{1em}$\displaystyle \boldsymbol{\theta}_{u,i}^{l} = \argmin_{\boldsymbol{\theta}_u} \mathcal{L}_{\rm PDE}(\boldsymbol{\theta}_u, \boldsymbol{\theta}_{m,i}^{l}) + \mathcal{L}_{\rm boundary}(\boldsymbol{\theta}_u)$
                \STATE Compute the function-space gradient $\mathbf{g}_{i}^{l} = \nabla_{\mathbf{m}_{i}^{l}}\mathcal{J}$ by the adjoint method
            \ENDFOR
            \STATE Compute the fParVI update vectors $\{\boldsymbol{\phi}(\mathbf{m}_{i}^{l})\}_{i=1}^{n_p}$ using Equation \ref{eqn:SVGD_vector}
            \FOR{$i=1,\ldots,n_p$}
                \STATE Update $\boldsymbol{\theta}_{m,i}^{l+1}$ from $\boldsymbol{\phi}(\mathbf{m}_{i}^{l})$ through the Jacobian mapping by Equation \ref{eqn:SVGD_update_improved2}
            \ENDFOR
        \ENDFOR
    \end{algorithmic}
\end{algorithm}

\begin{table}[htbp]
    \centering
    \caption{Comparison of MMD values between the semi-analytical solution and the posterior samples for different methods and two functional-prior settings. 
    The weight-space IID methods with and without RFF use $\sigma=0.7$ and $\sigma=0.3$ for both $l_\mathcal{GP}=0.075$ and $l_\mathcal{GP}=0.15$, respectively.}
    \label{tab:prior_method_comparison}
    \begin{tabular}{lcccccc}
        \toprule
        \shortstack{Functional \\ prior}& \shortstack{FPI-BPINN\\(2 layers)} & \shortstack{FPI-BPINN\\(3 layers)} & \shortstack{fParVI-PINN\\(2 layers)} & \shortstack{fParVI-PINN\\(3 layers)} & \shortstack{IID w/o RFF \\ (best case)} & \shortstack{IID w/ RFF \\ (best case)} \\
        \midrule
        $l_\mathcal{GP}=0.075$ & 0.123 & 0.141 & 0.0684 & 0.0773 & 0.216 & 0.160 \\
        $l_\mathcal{GP}=0.15$  & 0.144 & 0.149 & 0.0658 & 0.0681 & 0.211 & 0.151 \\
        \bottomrule
    \end{tabular}
\end{table}

\end{document}